\documentclass[fleqn,usenatbib]{mnras}

\usepackage{newtxtext,newtxmath}

\usepackage[T1]{fontenc}

\DeclareRobustCommand{\VAN}[3]{#2}
\let\VANthebibliography\thebibliography
\def\thebibliography{\DeclareRobustCommand{\VAN}[3]{##3}\VANthebibliography}

\usepackage{graphicx}	
\usepackage{amsmath}	

\title[Type 2 quasars]{The Quasar Feedback Survey: Revealing the Interplay of Jets, Winds \& Emission Line Gas in Type 2 Quasars with Radio Polarization}

\author[Silpa S. et al.]{
Silpa S.,$^{1}$\thanks{E-mail: silpa@ncra.tifr.res.in (SS)}
P. Kharb,$^{1}$
C. M. Harrison,$^{2}$
A. Girdhar,$^{3,4}$
D. Mukherjee,$^5$
V. Mainieri,$^3$
and M. E. Jarvis$^{3,4,6}$
\\
$^{1}$National Centre for Radio Astrophysics $-$ Tata Institute of Fundamental Research, S. P. Pune University Campus, Ganeshkhind, Pune 411007, India \\
$^{2}$School of Mathematics, Statistics and Physics, Newcastle University, Newcastle upon Tyne, NE1 7RU, UK\\
$^{3}$European Southern Observatory (ESO), Karl-Schwarzschild-Strasse 2, D-85748 Garching b. Muenchen, Germany\\
$^{4}$Ludwig Maximilian Universit{\"a}t, Professor-Huber-Platz 2, D-80539 Munich, Germany\\
$^5$Inter University Centre for Astronomy \& Astrophysics, Ganeshkhind, Pune 411007, India\\
$^{6}$Max-Planck Institut f\"ur Astrophysik, Karl-Schwarzschild-Str. 1, 85748 Garching, Germany
}

\date{Accepted XXX. Received YYY; in original form ZZZ}
\pubyear{2021}
\begin{document}
\label{firstpage}
\pagerange{\pageref{firstpage}--\pageref{lastpage}}
\maketitle

\begin{abstract}
We present results from a combined radio polarization and emission line study of five type 2 quasars at $z<0.2$ with the Karl G. Jansky Very Large Array (VLA) B-array at 5 GHz and Hubble Space Telescope (HST) [O~{\sc{iii}}] observations. These five sources are known to exhibit close association between radio structures and ionized gas morphology and kinematics. Four sources (J0945+1737, J1000+1242, J1356+1026 and J1430+1339) show polarization in the current data. J1010+1413 is the unpolarized source in our sample. We detect $0.5-1\%$ fractional polarization in the radio cores and a high fractional polarization ($10-30\%$) in the lobes of these sources. The morphological, spectral and polarization properties suggest a jet origin for radio emission in J0945+1737, J1000+1242, J1010+1413 and J1430+1339 whereas the current data cannot fully discern the origin of radio emission (jet or wind) in J1356+1026. An anti-correlation between various polarized knots in the radio and [O~{\sc{iii}}] emission is observed in our sources, similar to that observed in some radio-loud AGN in the literature. This suggests that the radio emission is likely to be depolarized by the emission-line gas. By modeling the depolarization effects, we estimate the size of the emission-line gas clouds to be $\sim(2.8\pm1.7)\times10^{-5}$ parsec and the amount of thermal material mixed with the synchrotron plasma to be $\sim(1.01\pm0.08)\times10^{6}$~M$_{\sun}$ in the lobe of J0945+1737 (which exhibits the most prominent polarization signature in its lobe). The current work demonstrates that the interplay of jets/winds and emission-line gas is most likely responsible for the nature of radio outflows in radio-quiet AGN. 
\end{abstract}

\begin{keywords}
galaxies: active -- quasars: general -- radio continuum: galaxies -- techniques: interferometric -- techniques: polarimetric 
\end{keywords}

\section{Introduction}
\label{intro}
Active galactic nuclei (AGN) are the energetic centres of galaxies and are powered by the release of gravitational potential energy as the matter accretes on to the supermassive black holes \citep[SMBHs; $10^6-10^9$~M$_\odot$; see review by][]{Rees84}. Quasars are extremely luminous AGN with the highest accretion rates. Types 1 and 2 are AGN subdivisions based on the presence or absence of broad emission lines \citep[with velocity widths of $\sim$1000~km~s$^{-1}$ in AGN optical spectra;][]{Antonucci93}. While quasars have been traditionally associated with type 1 spectra \citep{Peterson97}, type 2 quasars have been identified by \citet{Zakamska03,Zakamska04,Reyes08,Mullaney13}, based on their optical spectra. \citet{Norman02} detected one of the first X-ray selected high-redshift type 2 quasar.

AGN are thought to play a crucial role in the growth of galaxies by either suppressing or inducing star-formation \citep[e.g. ][]{AlexanderHickox12,Fabian12,Morganti17, Harrison17a}. They can either cause a `negative feedback' effect by expelling and heating the star-forming material via outflows and resulting in a shut-down of star-formation \citep[e.g. ][]{SilkRees98,Hopkins06,BoothSchaye10,KingPounds15,Costa18} or can cause a `positive feedback' effect by triggering local enhancements of star-formation \citep[e.g.][]{Ishibashi12,Silk13,Zubovas13,Lacy17,Fragile17,Gallagher19}. While the AGN feedback mechanism is believed to be fundamental in the galaxy formation and evolution processes from a theoretical point of view \citep[e.g.][]{Bower06,McCarthy10,Vogelsberger14,Schaye15,Choi18}, from an observational point of view there are many outstanding unsettled questions concerning the impact of AGN on the host galaxy interstellar medium (ISM) and star-forming properties \citep{VillarMartin16,Maiolino17,Kakkad17,YesufHo20,Scholtz21}.

Traditionally, feedback has been known to manifest in two flavours: quasar mode and maintenance mode \citep[e.g. see ][]{Croton09,Bower12}. The former is associated with radiative mode AGN, such as quasars that accrete at high Eddington ratios ($\geq0.01$) and release most of their energy in the form of radiation or accretion disk winds, and are predicted to result in rapid termination of star formation \citep[e.g. ][]{Faucher-GiguereQuataert12,KingPounds15,Harrison17b,Costa18}. Conversely, the latter mode is associated with low accretion rates and observations show that most of the energy is mechanical, traced via radio emission from jets, with the energy output able to regulate star formation \citep[e.g. ][]{McNamaraNulsen12,HardcastleCroston20}. However, recent works have challenged this traditional division and shown that both modes of feedback can be involved in the regulation of star formation \citep[e.g.][]{Churazov05,Ciotti10,Cielo18}.

The prevalence of galactic-scale multi-phase outflows in quasars is widely discussed in the literature \citep[e.g.][]{Liu13, Cicone14, Fiore17, BaeWoo18, Morganti18}. However, the relative contributions of different driving mechanisms for these outflows remain controversial, such as radiation pressure, jets, AGN winds and star-formation \citep[e.g.][]{Harrison18,WylezalekMorganti18,Hwang18,Jarvis19}. While radio jets are commonly regarded as the drivers of galactic-scale outflows in radio-loud (RL) AGN, the origin of radio emission in radio-quiet (RQ) AGN is poorly understood \citep{Panessa19}.   

Spatially-resolved radio observations and multi-phase spectroscopic observations are crucial for understanding the dominant processes giving rise to the observed radio emission in AGN and the various outflow mechanisms which transfer energy from the central engine to the host galaxies. Several works have shown strong correlations between the dynamics and kinematics of ionized gas outflows based on [O~{\sc{iii}}] emission-line profiles and radio luminosity \citep{Heckman81,Veilleux91,Mullaney13, ZakamskaGreene14}. 

Recent works have demonstrated the ability of radio polarimetric observations to distinguish between AGN jets/winds and starburst-driven winds based on the differences in their degrees of polarization, magnetic (B-) field structures and rotation measures (RM). \citet{Sebastian19a,Sebastian19b,Sebastian20} tentatively detect higher fractional polarization in Seyfert galaxies as compared to starburst galaxies. \citet{Silpa21a, Silpa21b} suggest poloidal B-field goemetry for jet/jet spine and toroidal B-field geometry for AGN wind, which could either be a magnetized accretion disc wind or the outer layers of a broadened jet (like a jet sheath) or a combination of both. Radio polarization in conjunction with spatially-resolved emission-line observations can provide useful diagnostics on the different types of media that co-exist and influence each other and also provide insights on fundamental questions pertaining to galactic outflows and feedback.

In this paper, we present results from our 5 GHz B-array (resolution$\sim$1 arcsec) Karl Jansky Very large Aray (VLA) polarization-sensitive observations combined with the archival Hubble Space Telescope (HST) [O~{\sc{iii}}] images for five type 2 RQ quasars belonging to the Quasar Feedback Survey sample \citep[QFeedS\footnote{https://blogs.ncl.ac.uk/quasarfeedbacksurvey/}; Figure 1;][]{Jarvis21}. Similar resolution VLA images from 2013-2014 and 2016-2017, along with higher resolution ($\sim$0.25 arcsec) ones, have been presented in \citet{Jarvis19}. However, these results provided no information on the polarization structures. Furthermore, our new data are more sensitive and also reveal details of the radio source structures better. Data from narrow-band HST filters around [O~{\sc{iii}}]$\lambda$5007 are used in the current work and are discussed in detail in the online supplementary material of \citet{Jarvis19}. We attempt here to understand the interplay of jets/winds and emission-line gas in these sources. We aim to establish the jet/wind origin of the radio emission from these sources and the impact of the local gaseous environments on the nature of radio outflows in them, through a joint radio polarization and emission-line study. We also aim to look for signatures of jet-medium interaction in these sources from the polarization perspective, as has been done for RL AGN in the literature.

Throughout this work, we adopt $\Lambda$CDM cosmology with $H_0$ = 73~km~s$^{-1}$~Mpc$^{-1}$, $\Omega_{m}$ = 0.27 and $\Omega_{v}$ = 0.73. Spectral index $\alpha$ is defined as $S_\nu\propto\nu^{\alpha}$, $S_\nu$ being the flux density at frequency $\nu$.

\section{Sample}
\label{sample}
\begin{figure}
\centerline{
\includegraphics[width=9cm]{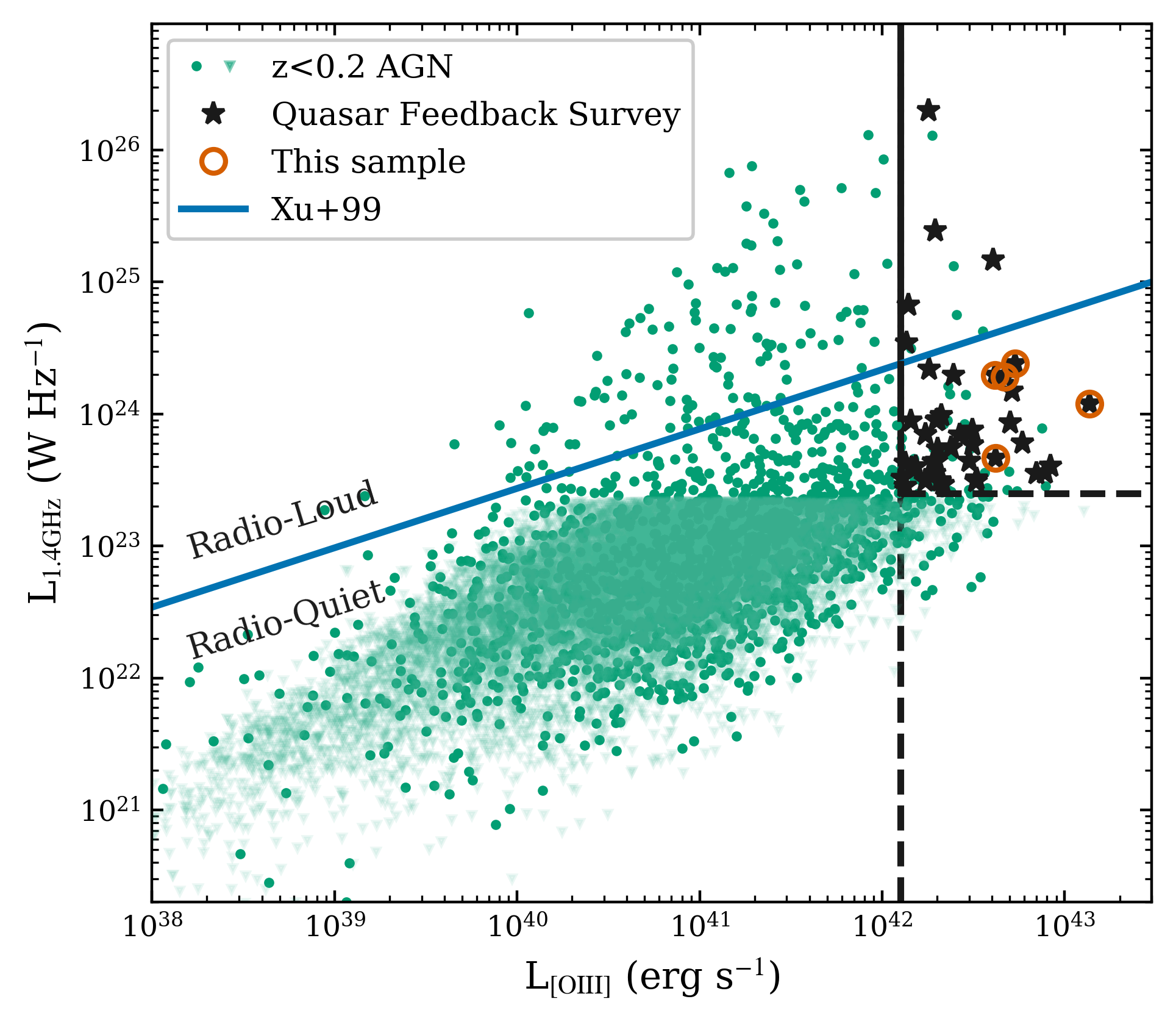}}
\caption{L$\mathrm{_{1.4GHz}}$ from  NVSS versus L$\mathrm{_{[O~III]}}$ for z $<$ 0.2 AGN from \citet{Mullaney13} plotted as green circles. Pale green triangles represent AGN with upper limits on their radio luminosity. The 42 QFeedS sources are plotted as black stars, and the 5 sources of current study highlighted in red circles. The blue line marks the division between 'radio-loud' and 'radio-quiet' AGN following the \citet{Xu99} definition. The black lines mark the selection criteria for QFeedS sources: L$\mathrm{_{1.4GHz}} > 10^{23.45}$ W~Hz$^{-1}$ and L$\mathrm{_{[O~III]}} > 10^{42.11}$ erg~s$^{-1}$.}
\label{fig1}
\end{figure}

Our sample consists of five type 2 RQ quasars selected from QFeedS, viz., J0945+1737, J1000+1242, J1010+1413, J1356+1026, J1430+1339. QFeedS comprises of 42 z $<0.2$ type 1 and type 2 AGN with quasar-like luminosities (L$_\mathrm{AGN} \geq 10^{45}~ \mathrm{erg~s}^{-1}$) spectroscopically-selected from \citet{Mullaney13}. These sources are also selected to have quasar-level [O~{\sc{iii}}] luminosities (L$\mathrm{_{[O~III]}} > 10^{42.11}$ erg~s$^{-1}$) and radio luminosities above the NRAO VLA Sky
Survey \citep[NVSS;][]{Condon98} detection limit (L$\mathrm{_{1.4GHz}} > 10^{23.45}$ W~Hz$^{-1}$; see Figure\ref{fig1}). QFeedS is designed to undertake a systematic study of the origin of radio emission in quasars, dominant drivers of galactic outflows, measurements of mass and energy content of the outflows and impact of quasars on their host galaxies, using spatially-resolved multi-wavelength observations. The description of the full sample and survey properties is provided in \citet{Jarvis21} while several pilot studies have been carried out on subsets of the QFeedS sources \citep{Harrison15, Lansbury18, Jarvis19, Jarvis20, Jarvis21, Girdhar22}.

The 5 sources in our sample are selected from the initial sample of 10 type 2 RQ quasars from \citet{Jarvis19}. These sources have been classified as RQ based on the definition of \citet[][shown in Figure~\ref{fig1}]{Xu99}; radio luminosities are higher by a factor of 10$^3$-10$^4$ for the RL sources compared to RQ ones in the 5~GHz radio luminosity versus [O~{\sc{iii}}]$\lambda$5007 line luminosity plots. These sources are known to host powerful and galactic scale ionized gas outflows (L$_\mathrm{[O~III]}>10^{42}\mathrm{~erg~s^{-1}}$; $\mathrm{FWHM_{[O~III]}\approx 800 - 1800~km~s^{-1}}$), kpc-scale radio structures likely dominated by jet emission and signatures of jet-gas interactions on galactic scales such as increased turbulence of the gas and outflowing bubbles, based on spatially resolved radio observations and Integral Field Spectroscopy (IFS) data \citep{Harrison14, Harrison15, Lansbury18, Jarvis19}. The 5 sources are specifically chosen due to the prevalence of prominent radio structures, such as jets, lobes and bubbles in \citet{Jarvis19} and a close association between the radio structures and the ionized gas kinematics and dynamics. Therefore, these sources serve as potential candidates to understand the interplay of jets/winds and emission-line gas using a combined radio polarization and emission-line study.

\section{Data Analysis}
\label{data_analysis}
The VLA 5 GHz B-array observations ($\theta\sim1$~arcsec) of all 5 sources were carried out on 6 and 7 July, 2020 under the Project ID: 20A-176. 3C286 was used as both flux and polarized calibrator whereas OQ208 was used as the unpolarized calibrator. The phase calibrators for individual sources are listed in Table~\ref{table1}. The basic properties of individual sources, and the details of their VLA images are summarized in Table~\ref{table1}. Basic calibration and editing of the VLA data was carried out using the $\tt{CASA}$ calibration pipeline for VLA data reduction. This was followed by manual polarization calibration steps, which involved (i) setting the polarization model for the polarized calibrator, (ii) solving for the cross-hand (RL, LR) delays arising from the residual delay difference between R and L signals, (iii) solving for the instrumental polarization (`D-terms') arising from the imperfections and non-orthogonality of the antenna feeds, and (iv) solving for the R-L phase offset arising from the difference between the R and L gain phases for the reference antenna \citep[see][for details]{Silpa21a}. The average D-term amplitude turned out to be typically $\sim 5$\%. 

The visibility data for individual sources were extracted from the calibrated multi-source dataset and imaged for Stokes I using the multiterm-multifrequency synthesis \citep[MT-MFS;][]{RauCornwell11} algorithm of $\tt{TCLEAN}$ task in $\tt{CASA}$. Three rounds of phase-only self-calibration and one round of amplitude and phase self-calibration were carried out. Stokes Q and U images were created from the last self-calibrated visibility data using the same imaging parameters as for the Stokes I image but for fewer number of iterations. The Gaussian-fitting $\tt{AIPS}$ task $\tt{JMFIT}$ was used to obtain the flux densities of the compact components like the core, whereas the $\tt{AIPS}$ task $\tt{TVSTAT}$ was used for extended emission. The distances reported in this paper were estimated using the $\tt{AIPS}$ task $\tt{TVDIST}$.

\subsection{VLA polarization imaging}
\label{poln_imaging}
The Stokes Q and U images were combined using the $\tt{AIPS}$ task $\tt{COMB}$ with $\tt{opcode=POLC}$ (for Ricean bias correction) to create linear polarized intensity image ($P=\sqrt{Q^2+U^2}$) and with $\tt{opcode=POLA}$ to create polarization angle image ($\chi$ = 0.5 tan$^{-1}$(U/Q)). Regions with intensity values below $\sim$3 times the rms noise and angle values with errors $>$ $10^\circ$ were blanked while making the {\tt PPOL} and {\tt PANG} images respectively. The fractional polarization ($FP= P/I; FPOL$) image was created from {\tt PPOL} and Stokes I images using the task $\tt{COMB}$ with {\tt opcode = DIV} while blanking regions with fractional polarization errors $>$10\%.

\subsubsection{VLA sub-band polarization imaging for J0945+1737}
\label{inband_poln_imaging}
\begin{figure*}
\centerline{
\includegraphics[width=10.2cm]{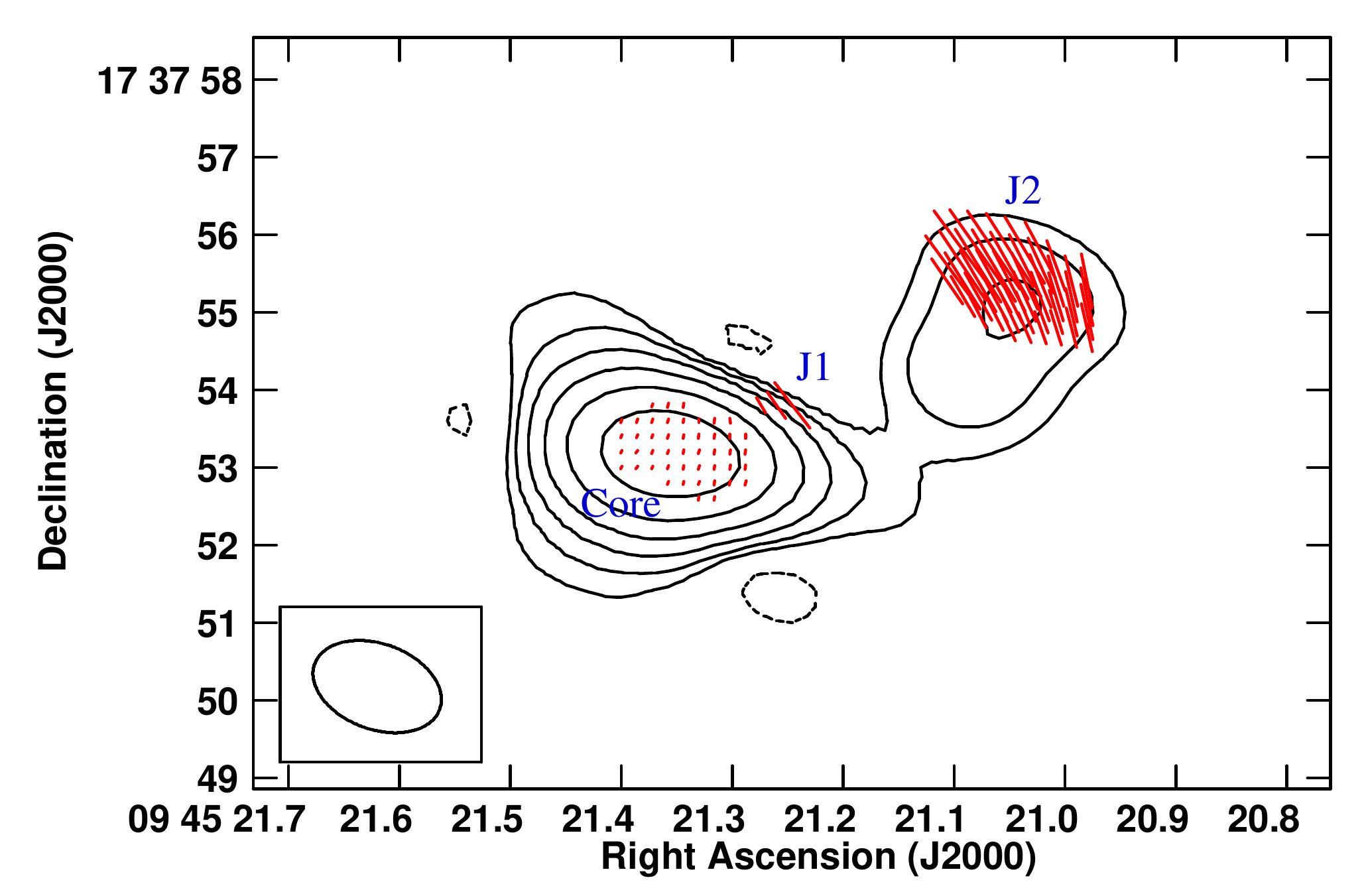}
\includegraphics[width=10cm]{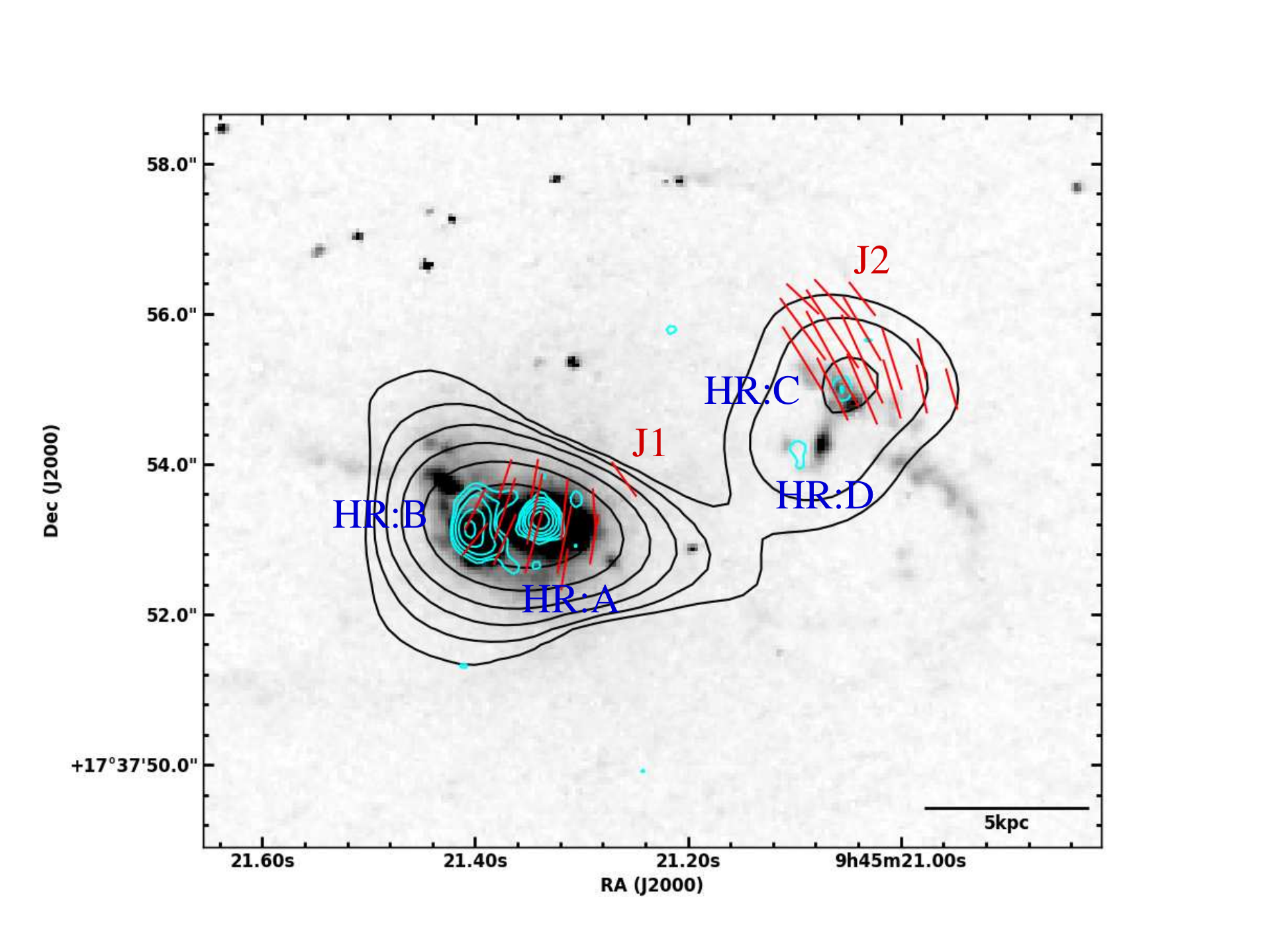}}
\centerline{
\includegraphics[width=7.5cm,trim=80 150 30 150]{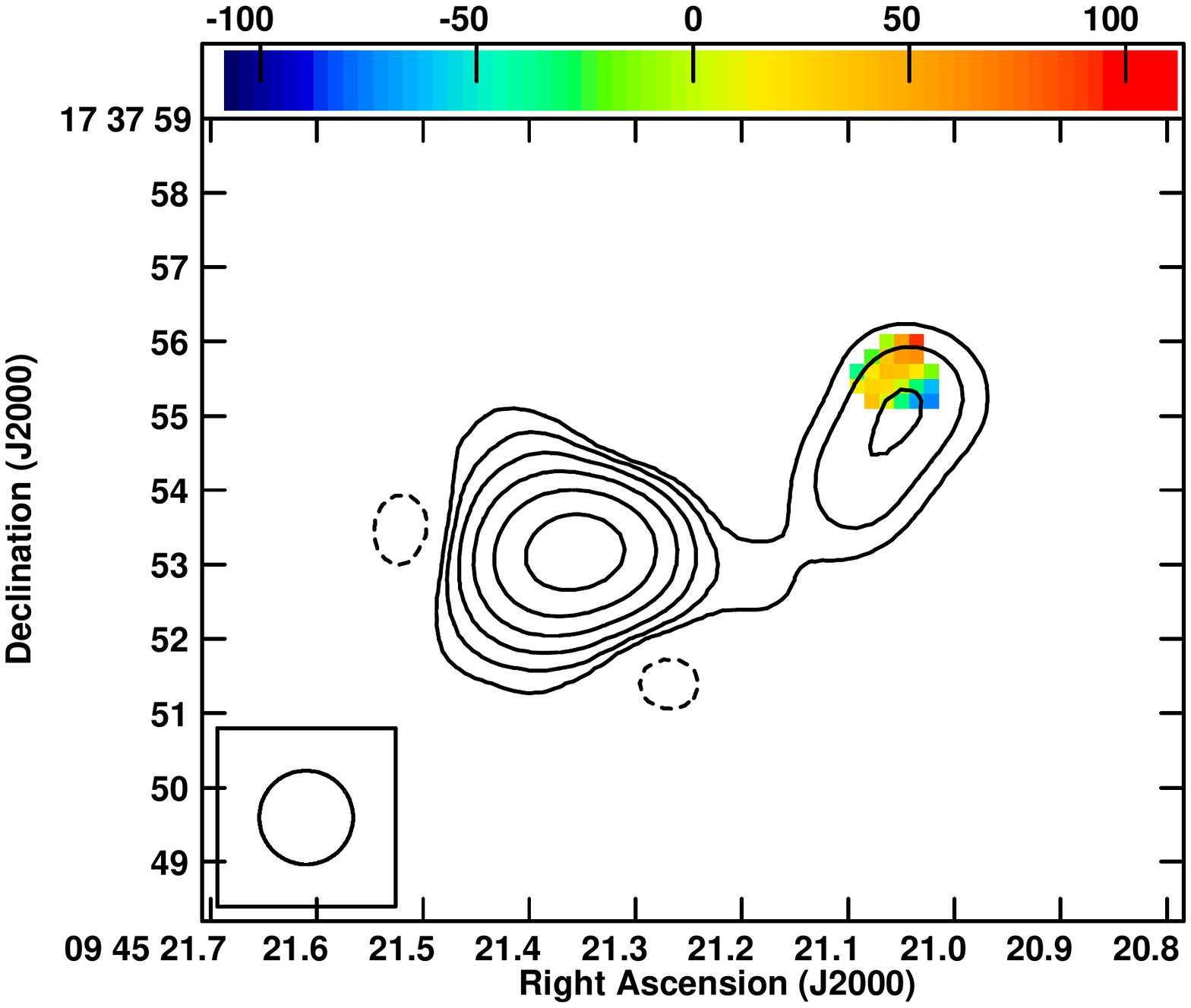}}
\caption{VLA 5 GHz B-array (Beam: 1.71 $\times$ 1.10 arcsec, 70.01$\degr$) total intensity contours in black for J0945+1737 with (top left) red superimposed electric polarization vectors, (top right) gray-scale HST [O~{\sc{iii}}] image, and (bottom) color in-band ($\sim$ 4$-$6 GHz) RM image. In the top left panel, the length of the vectors are proportional to fractional polarization with 1$\arcsec$ corresponding to 10\% fractional polarization. In the top right panel, the VLA 6 GHz A-array \citep[$\theta\sim$ 0.25 arcsec;][]{Jarvis19} total intensity contours are in cyan with the polarization vectors proportional to polarised intensity. The color scale ranges $-$110 to 110 rad~m$^{-2}$ in the bottom panel. The peak contour surface brightness is 9.4 mJy beam$^{-1}$ and the levels are 0.17 $\times$ (-1, 1, 2, 4, 8, 16, 32, 64, 128, 256, 512) mJy beam$^{-1}$ for the VLA 5 GHz B-array image in the top panels. The contour levels are 0.019 $\times$ (4, 8, 16, 32, 64, 128) mJy beam$^{-1}$ for the VLA 6 GHz A-array image. The bottom panel image is made with a 1.26 $\times$ 1.26 arcsec beam with the peak contour surface brightness of 8.7 mJy beam$^{-1}$ and the levels of 0.18 $\times$ (-1, 1, 2, 4, 8, 16, 32, 64, 128, 256, 512) mJy beam$^{-1}$. The J and CJ labels correspond to the polarization features and HR labels correspond to the $\sim$0.25$\arcsec$ features in all figures. North is up and east is to the left for all figures.}
\label{fig2}
\end{figure*}

As the polarization signature in the lobe was most prominent ($>7\%$ in 4 individual sub-band images) in the case of J0945+1737 (see Figure~\ref{fig2}), we chose this source for modeling depolarization as discussed in Section~\ref{depoln} ahead. The calibrated VLA 5 GHz data with 16 spectral windows spanning a total bandwidth of $\sim$2 GHz were divided into 4 sub-bands, each spanning a bandwidth of 512 MHz. The Stokes I, Q and U images convolved with identical circular beam (1.26 $\times$ 1.26 arcsec) were created from individual sub-band datasets using the $\tt{CASA}$ task $\tt{TCLEAN}$. The {\tt PPOL}, {\tt PANG} and {\tt FPOL} images were created for individual sub-band datasets.

In order to make an in-band ($\sim$4$-$6 GHz) RM image for J0945+1737, Stokes QU image cubes were created from individual sub-band datasets using the $\tt{CASA}$ task $\tt{TCLEAN}$. These images were convolved with identical circular beam (1.26 $\times$ 1.26 arcsec). The task $\tt{RMFIT}$ in $\tt{CASA}$ was used to create the in-band RM image from these image cubes. Pixels with RM error greater than 155~rad~m$^{-2}$ were blanked. The reported RM error was obtained from the RM noise image created using $\tt{RMFIT}$.

\begin{table*}
\begin{center}
\caption{Basic properties of the sample sources and summary of their radio images}
\label{table1}
{\begin{tabular}{ccccccccl}
\hline
Quasar & R.A. (J2000.0) & Decl. (J2000.0) & Redshift & log(L$\mathrm{_{[O~III]}}$)$^*$ & log(L$\mathrm{_{1.4GHz}}$)$^{**}$ & Phase cal & $\it{rms}$ noise & Beam, PA\\
& (hh:mm:ss.ss)  & (+/-dd:mm:ss.s) & & (erg s$^{-1}$) & (W Hz$^{-1}$) & & ($\mu$Jy~beam$^{-1}$) & (arcsec $\times$ arcsec, $\degr$) \\
\hline
J0945+1737 & 09:45:21.33 & +17:37:53.2 & 0.128 & 42.67 & 24.3 & J0956+2515 & 7 & 1.71 $\times$ 1.10, 70.01 \\
J1000+1242 & 10:00:13.14 & +12:42:26.2 & 0.148 & 42.62 & 24.3 & J1016+2037 & 8 & 2.50 $\times$ 1.09, 61.14 \\
J1010+1413 & 10:10:22.95 & +14:13:00.9 & 0.199 & 43.14 & 24.1 & J1016+2037 & 10 & 2.41 $\times$ 1.16, 61.50 \\
J1356+1026 & 13:56:46.10 & +10:26:09.0 & 0.123 & 42.73 & 24.4 & J1347+1217 & 7 & 1.40 $\times$ 1.08, $-$48.77 \\
J1430+1339 & 14:30:29.88 & +13:39:12.0 & 0.085 & 42.62 & 23.7 & J1415+1320 & 7 & 1.31 $\times$ 1.19, $-$52.02 \\
\hline
\end{tabular}}

$^*$ Total observed [O~{\sc{iii}}]$\lambda$5007 luminosities estimated using the fluxes from \citet{Mullaney13}, as reported in \citet{Jarvis21} (see their table 1).\\
$^{**}$ Rest-frame 1.4 GHz radio luminosities from NVSS assuming $S_\nu\propto\nu^{\alpha}$ and $\alpha$=$-$0.7, as reported in \citet{Jarvis21} (see their table 1).
\end{center}
\end{table*}

\section{Results}
\label{results}
We present VLA 5 GHz B-array ($\theta\sim1$~arcsec) total intensity images in black contours superimposed with polarization electric ($\chi$) vectors in red whose lengths are proportional to fractional polarization, for individual sources in the top left panel of Figure~\ref{fig2}, top panel of Figure~\ref{fig3}, and left panels of Figs Figures~\ref{fig4}-\ref{fig6}. 

We detect polarization in all sources, except J1010+1413 with the current data. We label the polarization structures in the core, jet/lobe and counter-jet/lobe components as Core, Jn and CJn (where n=1,2,3,4) respectively. The polarized flux density, fractional polarization and polarization angle values for different radio components of individual sources are presented in Table~\ref{table2}. The cores exhibit about $0.5-1$\% fractional polarization, which is typical even for RL AGN, whereas high fractional polarization ($10-30\%$) is revealed in the lobes. The convention following the synchrotron theory suggests that the inferred B-field vectors are perpendicular to the $\chi$-vectors for optically thin regions, and parallel for optically thick regions. We adopt this convention throughout the paper for interpreting our polarization results. 

We present HST [O~{\sc{iii}}] images in gray-scale superimposed with VLA 5~GHz B-array total intensity contours in black, VLA 6~GHz A-array \citep[$\theta\sim0.25$~arcsec; from][]{Jarvis19} total intensity contours in cyan and electric polarization vectors in red, whose lengths are proportional to polarised intensity for individual sources in the top right panel of Figure~\ref{fig2}, bottom panel of Figure~\ref{fig3} and right panels of Figures~\ref{fig4}-\ref{fig6}. We adopt the nomenclature (i.e., `HR' labels) used by \citet{Jarvis19} for the VLA 6 GHz A-array images.

We estimate jet kinetic power ({\it P}$_\mathrm{jet}$) of individual sources using the following empirical relation obtained for a sample of low-luminosity radio galaxies by \citet{MerloniHeinz07}:
\begin{equation}
\mathrm{log({\it P}_{jet}) = 0.81~log({\it L}_{5GHz}) + 11.9},
\end{equation}
where $\mathrm{L_{5GHz}}$ is the 5 GHz radio core luminosity. The values of 5 GHz flux density for different radio components ({\it S}$_\mathrm{5GHz}$), {\it L}$_\mathrm{5GHz}$ and {\it P}$_\mathrm{jet}$ for individual sources are tabulated in Table~\ref{table3}. {\it P}$_\mathrm{jet}$ of the order of 10$^{43}-10^{44}$ erg~s$^{-1}$ is suggestive of low-power jets in these sources.

We present morphology, spectral index properties, polarization properties and relation between polarization \& ionized gas for individual sources in Sections~\ref{J0945} to \ref{J1430}. We discuss these results in Section~\ref{discuss} and summarize them in Section~\ref{conclusions}. 

\subsection{J0945+1737}
\label{J0945}
We detect a radio core, a north-western (NW) lobe and a curved jet connecting the two (black contours in Figure~\ref{fig2}). The core and lobe have also been detected by \citet{Jarvis19} and \citet{VillarMartin21} in their $\sim1$~arcsec VLA images. The $\sim0.25$~arcsec image of \citet{Jarvis19} reveals a core (HR:A), a bent-jet structure to the east (HR:B) and two hotspots (HR:C and HR:D) embedded inside the $\sim1$~arcsec lobe (cyan contours in Figure~\ref{fig2}, top right panel). 

We detect three distinctly polarized regions in J0945+1737 (see Figure~\ref{fig2}, top panels). The core and a region close to the core (J1) are polarized. Then a region at the end of the lobe (J2) is highly polarized. Interestingly, HR:C is coincident with the J2 region (see Figure~\ref{fig2}, top right panel). The inferred B-field vectors at J2 are neither parallel nor perpendicular to the local direction of the jet but appear more oblique. We also find more polarization (viz. J1 and J2) on the opposite side of the $\sim0.25$~arcsec jet (HR:B). The inferred B-field in the optically thin core as indicated from its steep spectrum (see Table~\ref{table3} and Appendix~\ref{appendixA}) stays roughly aligned with the direction of the $\sim1$~arcsec curved jet. 

A spatial overlap between radio emission and [O~{\sc{iii}}] emission has  been identified in this source by \citet{Jarvis19}. We find an anti-correlation between the polarized knots J1 and J2, and [O~{\sc{iii}}] emission (see Figure~\ref{fig2}, top right panel). These knots are located in regions with a dearth of [O~{\sc{iii}}] emission, except for a small region of overlap with the emission-line gas at J2 that is co-spatial with HR:C. We note that the region between HR:C and HR:D has [O~{\sc{iii}}] emission but no polarization. Of the four sources showing polarization, J0945+1737 is the only one with significant polarization visible in sub-band images. In Section~\ref{depoln}, we discuss different depolarization mechanisms in this source.

\subsection{J1000+1242}
\label{J1000}
\begin{figure*}
\centerline{
\includegraphics[width=10.5cm]{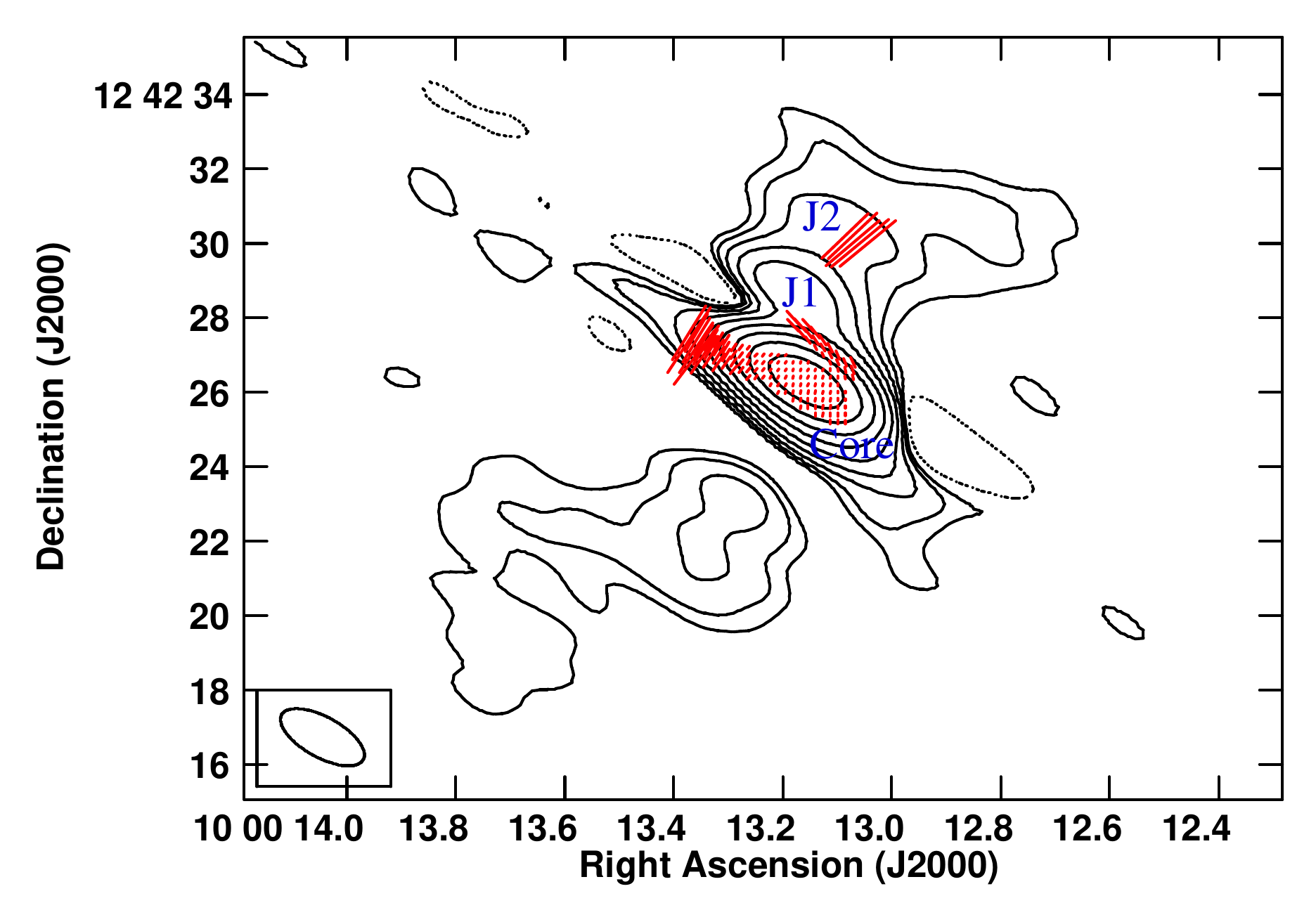}}
\centerline{
\includegraphics[width=15cm]{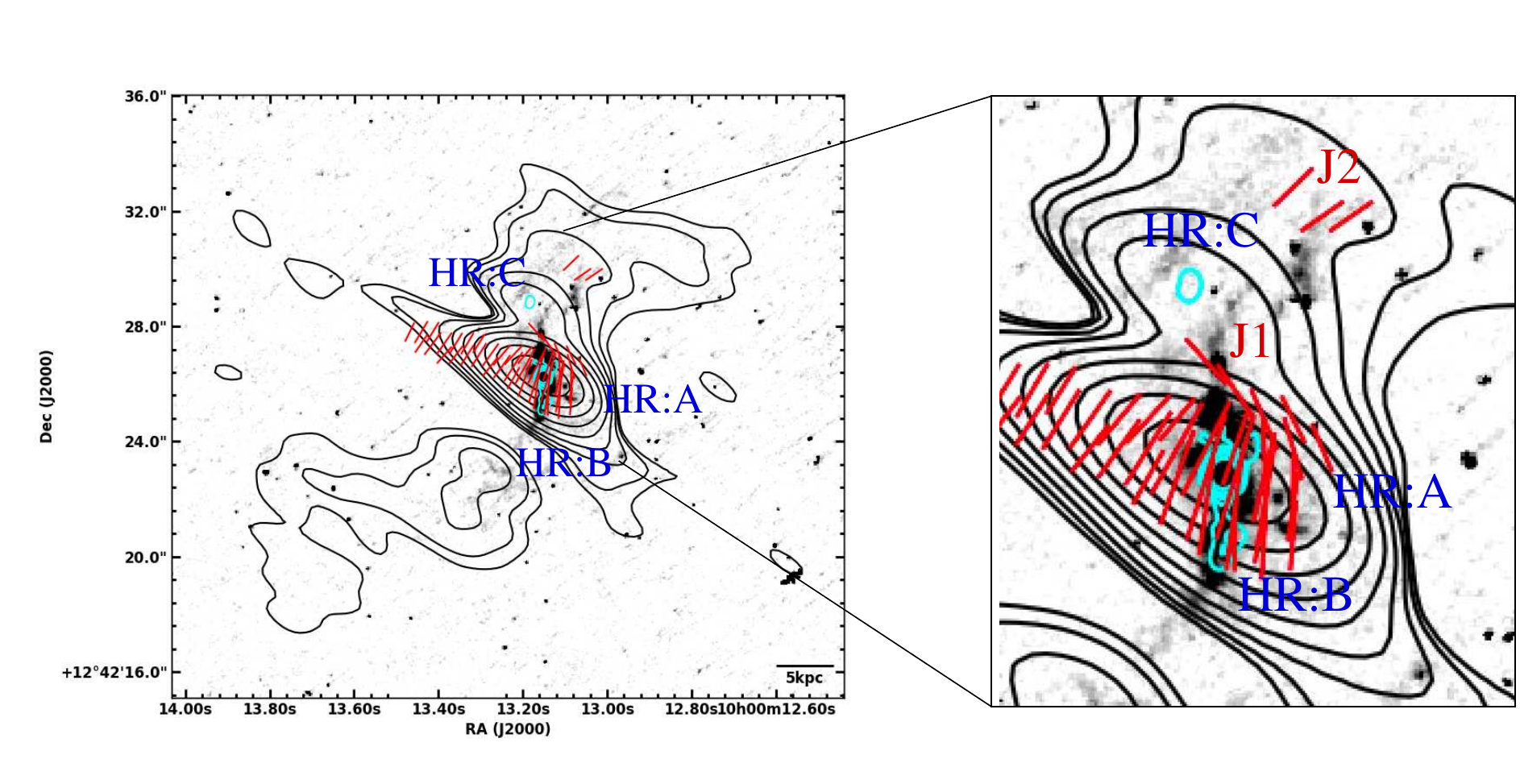}}
\caption{VLA 5 GHz B-array (Beam: 2.50 $\times$ 1.09 arcsec, 61.14$\degr$) total intensity contours in black for J1000+1242 with (top) red superimposed electric polarization vectors, and (bottom left) gray-scale HST [O~{\sc{iii}}] image with a zoom-in of its central region in the bottom right. In the top panel, the length of the vectors are proportional to fractional polarization with 1$\arcsec$ corresponding to 8.3\% fractional polarization. In the bottom left panel, the VLA 6 GHz A-array total intensity contours are in cyan with the polarization vectors proportional to polarised intensity. The peak contour surface brightness is 12 mJy beam$^{-1}$ and the levels are 0.0265 $\times$ (-1, 1, 2, 4, 8, 16, 32, 64, 128, 256, 512) mJy beam$^{-1}$ for the VLA 5 GHz B-array image. The contour levels are 0.023 $\times$ (8, 16, 32, 64, 128) mJy beam$^{-1}$ for the VLA 6 GHz A-array image.}
\label{fig3}
\end{figure*}

We detect a radio core and a deflected jet giving rise to the diffuse lobes in the south-east (SE) and north-west (black contours in Figure~\ref{fig3}). These morphological features have been detected in the $\sim1$~arcsec images of \citet{Jarvis19} and \citet{VillarMartin21}, with the latter also detecting additional faint knots extending beyond the SE lobe towards the east. The $\sim0.25$~arcsec image of \citet{Jarvis19} reveals a core-jet structure (HR:A and HR:B) extending to the south and a hotspot (HR:C) embedded in the NW lobe (cyan contours in Figure~\ref{fig3}, bottom panels).

We find three distinctly polarized regions in this source (see Figure~\ref{fig3}): the core, a region close to the core (J1) and a region in the NW lobe (J2). The inferred B-fields in the optically thin core (see Table~\ref{table3} and Appendix~\ref{appendixA}) are transverse to the local jet direction (i.e. heading north-west). The direction of the inferred B-fields in the J1 component are offset from the local jet direction (which is identified by the line joining HR:A to HR:C). The inferred B-fields at position J2 are transverse to the local jet direction. The polarization is more prominent (viz. J1 and J2) on the opposite side of the $\sim0.25\arcsec$ jet (HR:B). We also note that HR:C is located between J1 and J2 components and the region with compressed B-fields (i.e. J2) is just a little to the north-west of HR:C. The polarisation feature in the north-east of the core is an imaging artefact and has large errors, but has been included in the figure in order to capture the regions with fainter polarized emission elsewhere.

A spatial correlation between radio emission and [O~{\sc{iii}}] emission in this source has been reported by \citet{Jarvis19}. We note that both phases possibly have an S-shaped symmetry (see Figure~\ref{fig3}, bottom panels). We also find that the polarized knots J1 and J2 are located in regions deficit of [O~{\sc{iii}}] emission. This anti-correlation may suggest that the emission-line gas is depolarizing the radio emission. The filamentary nature of [O~{\sc{iii}}] emission is also worth noting. There is a brightened [O~{\sc{iii}}] cloud and an interesting line splitting of [O~{\sc{iii}}] kinematics at the location of HR:C \citep{Jarvis19}. It seems that the jet might have collided with a gas cloud in its path that may have been there from a merger remnant (or a possible historic outflow). This might have have deflected and decollimated the jet, resulting in the distorted lobes. The collision may have caused a shock in this gas cloud and made it luminous in [O~{\sc{iii}}]. However, the offset between the locations of HR:C and J2 might also mean that the [O~{\sc{iii}}] is gas that a shock has passed through and is now cooling. There is also a spatial coincidence between the $\sim0.25\arcsec$ jet (HR:B) and the emission-line filament extending to the south. Moreover, the inferred B-fields in this region are transverse to the direction of the HR:B jet (see Figure~\ref{fig3}, bottom right panel), suggestive of shocks. Future work using IFU data on this source could investigate this scenario further.

\subsection{J1010+1413}
\label{J1010}
\begin{figure*}
\centerline{
\includegraphics[width=9cm, trim=0 170 0 150]{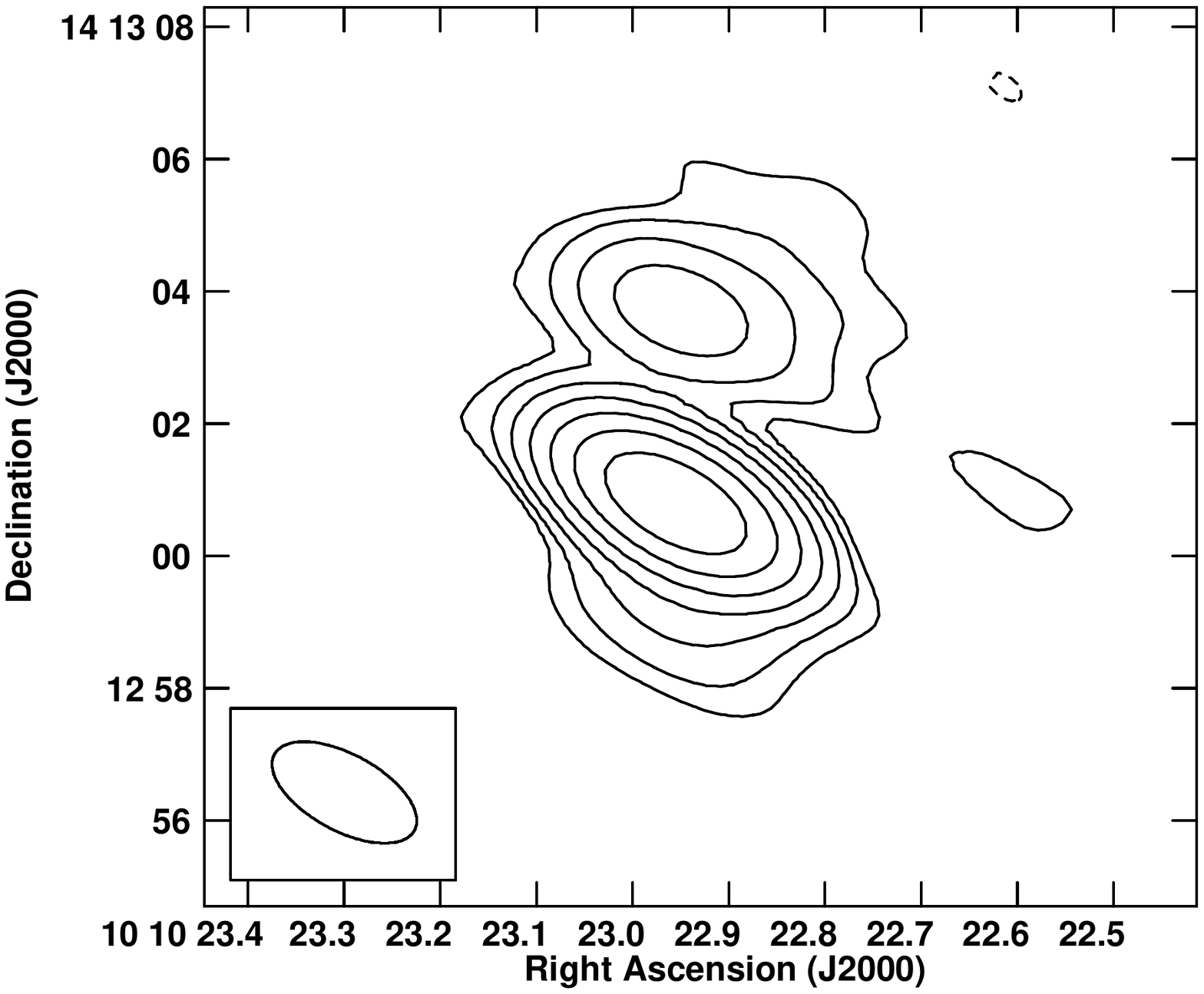}
\includegraphics[width=9cm, trim=0 20 70 150]{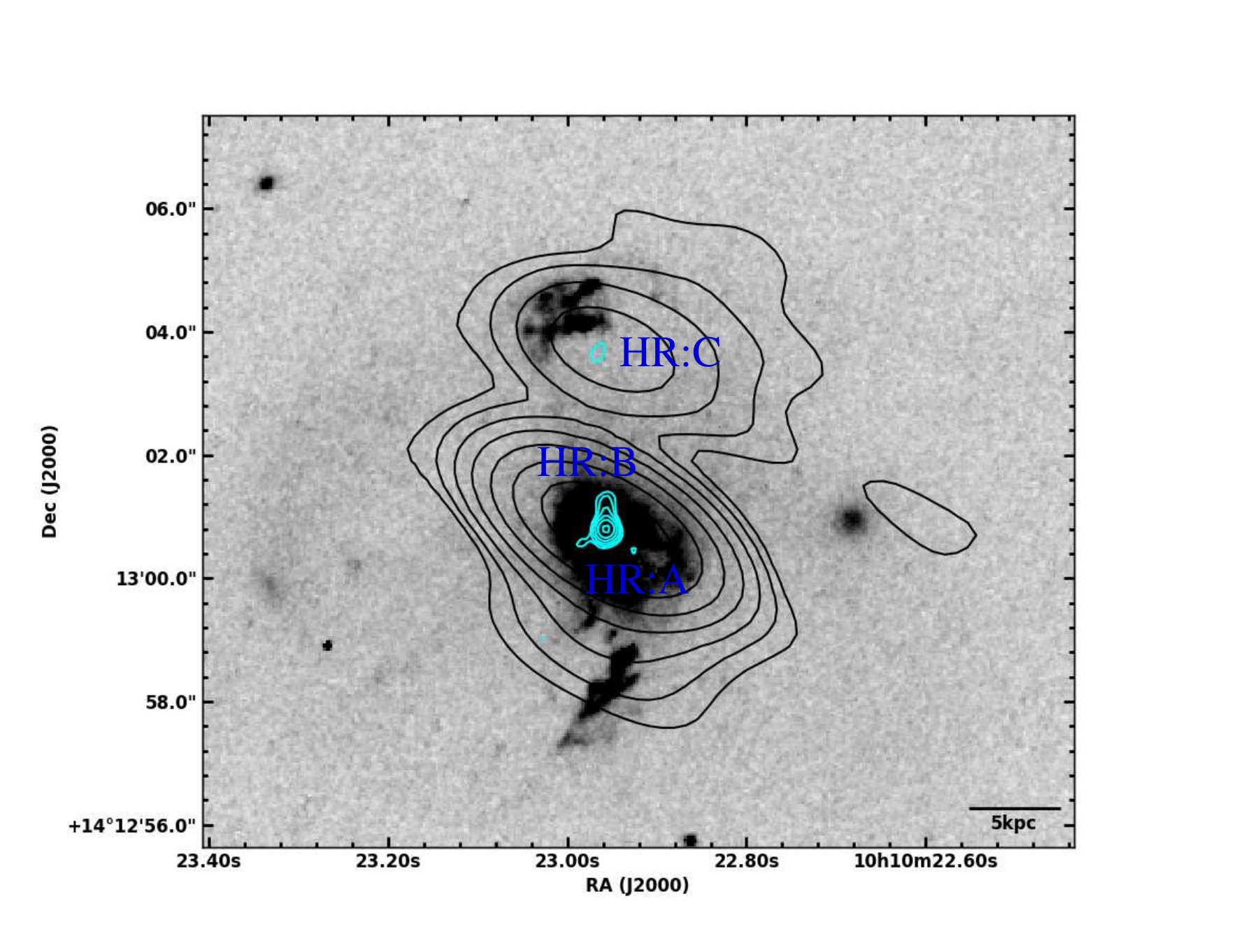}}
\caption{VLA 5 GHz B-array (Beam: 2.41 $\times$ 1.16 arcsec, 61.50$\degr$) total intensity contours in black for J1010+1413 on left, and with gray-scale HST [O~{\sc{iii}}] image on right. Since this is an unpolarized source, the polarization vectors are absent. In the right panel, the VLA 6 GHz A-array total intensity contours are in cyan. The peak contour surface brightness is 3.8 mJy beam$^{-1}$ and the levels are 0.031 $\times$ (-1, 1, 2, 4, 8, 16, 32, 64, 128, 256, 512) mJy beam$^{-1}$ for the VLA 5 GHz B-array image. The contour levels are 0.016 $\times$ (4, 8, 16, 32, 64, 128) mJy beam$^{-1}$ for the VLA 6 GHz A-array image.}
\label{fig4}
\end{figure*}

This source has two bubble-like lobes in the north-south direction (black contours in Figure~\ref{fig4}), which are indicative of those seen in Seyfert galaxies \citep[see e.g., ][]{Kharb06,HotaSaikia06}. The southern lobe is fainter and less extended than the northern one. The lobe-core-lobe morphology is also detected in the $\sim1\arcsec$ image of \citet{Jarvis19} whereas their $\sim0.25\arcsec$ image (cyan contours in Figure~\ref{fig4}, right panel) reveals possibly a core-jet structure oriented north-south (HR:A and HR:B), and a hotspot (HR:C) embedded in the more diffuse northern lobe.

As already mentioned, this is the only unpolarized source in our sample (hence the missing $\chi$ vectors in Figure~\ref{fig4}). This could either be attributed to the lack of sensitivity of our observations or a strong depolarization effect (see Section~\ref{caveats} for more details). 

The [O~{\sc{iii}}] emission seems to have an S-shaped symmetry along the radio lobes (see Figure~\ref{fig4}, right panel). We find an offset between [O~{\sc{iii}}] emission and peak radio emission in the northern lobe. This may suggest that either the emission-line clouds do not penetrate deeply into the lobe or that they evaporate quickly once inside \citep[e.g.][]{vanBreugel84}. It could also be that as the radio outflow propagates through the ambient medium, it entrains and compresses the local emission-line gas, which then rapidly cools and produces luminous line emission along the boundary of the radio structure. 

Such an offset is not seen in the southern lobe. Overall, it appears that the emission-line clouds along the boundary of the northern lobe may act like a foreground Faraday screen that depolarize the lobe emission while the mixing of the thermal ionized gas with the synchrotron plasma inside the southern lobe may result in the internal Faraday depolarization of the lobe emission.
 
\subsection{J1356+1026}
\label{J1356}
\begin{figure*}
\centerline{
\includegraphics[width=8.8cm,trim=0 0 0 0]{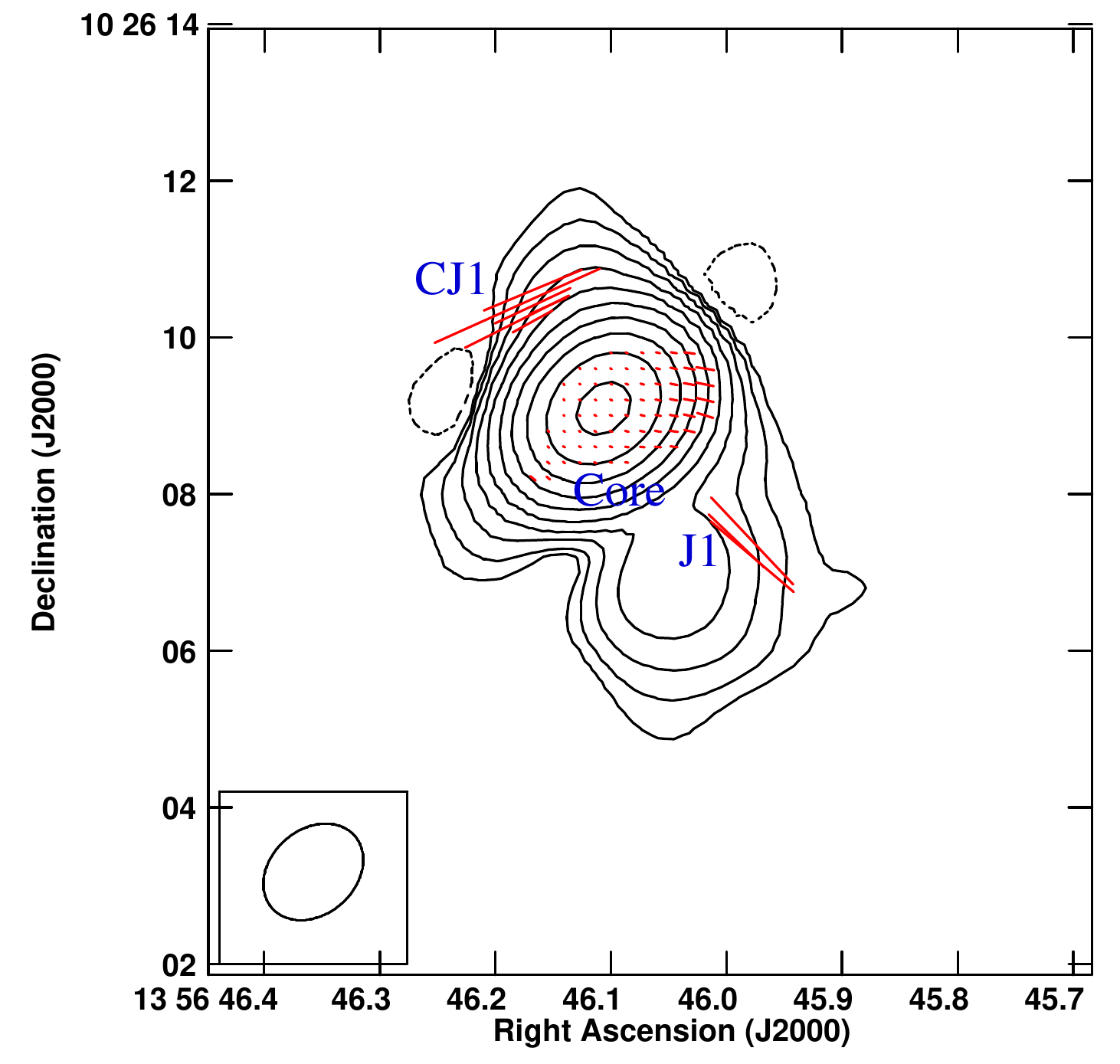}
\includegraphics[width=10.8cm,trim=0 20 0 0]{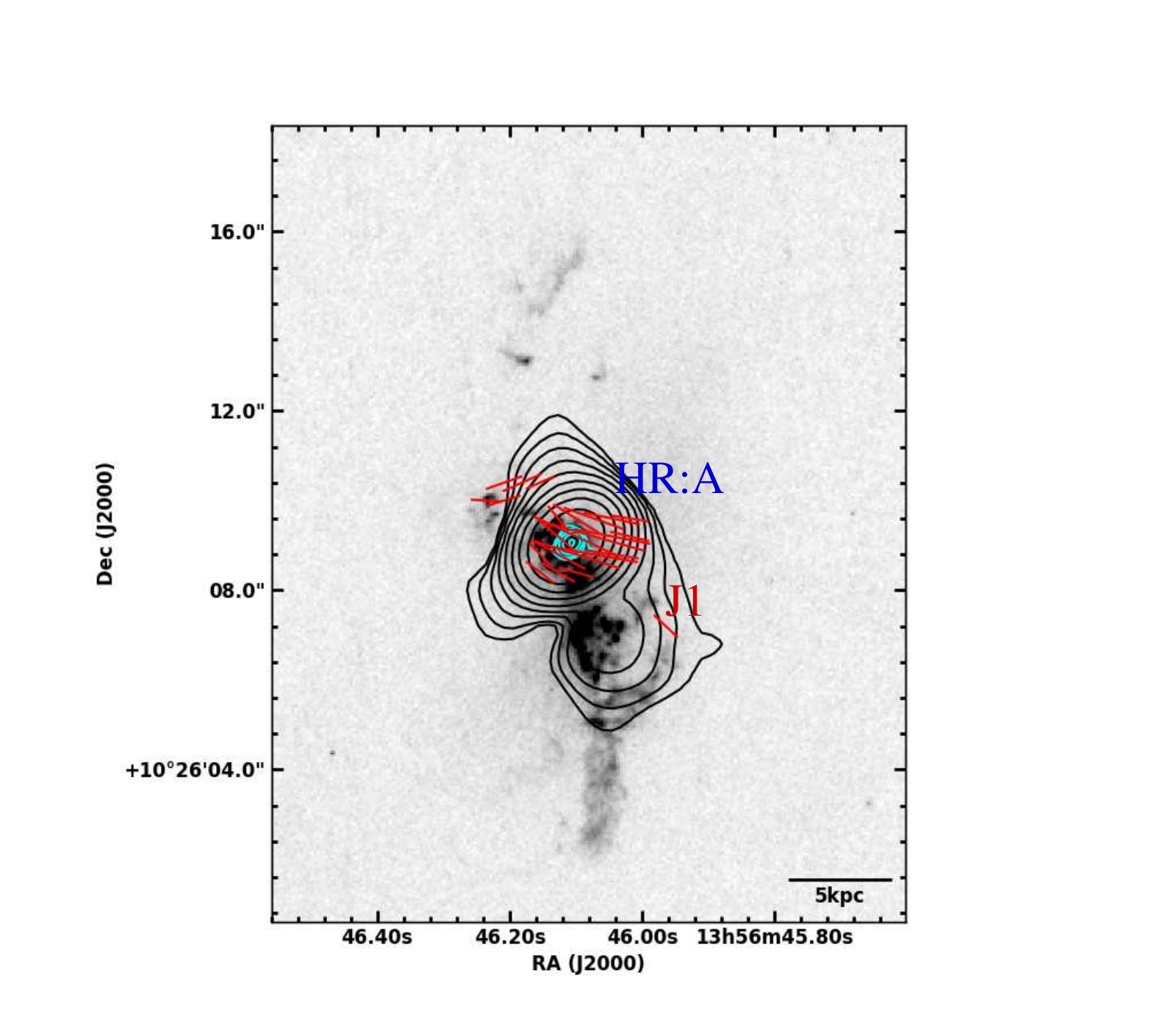}}
\caption{VLA 5 GHz B-array (Beam: 1.40 $\times$ 1.08 arcsec, $-48.77\degr$) total intensity contours in black for J1356+1026 with (left) red superimposed electric polarization vectors, (right) gray-scale HST [O~{\sc{iii}}] image. In the left panel, the length of the vectors are proportional to fractional polarization with 1$\arcsec$ corresponding to 12.5\% fractional polarization. In the right panel, the VLA 6 GHz A-array total intensity contours are in cyan with the polarization vectors proportional to polarised intensity. The peak contour surface brightness is 18~mJy~beam$^{-1}$ and the levels are 0.030 $\times$ (1, 2, 4, 8, 16, 32, 64, 128, 256, 512)~mJy~beam$^{-1}$ for the VLA 5 GHz B-array image. The contour levels are 0.1 $\times$ (8, 16, 32, 64, 128)~mJy~beam$^{-1}$ for the VLA 6 GHz A-array image.}
\label{fig5}
\end{figure*}

This source reveals a radio core and a diffuse southern lobe that curves slightly towards the east (black contours in Figure~\ref{fig5}). These structures have also been detected in the $\sim1$~arcsec images of \citet{Jarvis19} and \citet{VillarMartin21}. The $\sim0.25$~arcsec image of \citet{Jarvis19} detects only the core (HR:A) (cyan contours in Figure~\ref{fig5}, right panel). The $\sim4$~arcsec image of \citet{VillarMartin21} reveals faint large-scale emission extending to $\sim160$~kpc, resembling emission from a pair of outer radio lobes at the same position angle (PA) as that of the inner lobe/jet.

We detect three distinctly polarized regions in this source (see Figure~\ref{fig5}): the core, a region in the southern lobe (J1) and a region to the north-east of the core (CJ1). For an optically thin core (see Table~\ref{table3} and Appendix~\ref{appendixA}), the inferred B-fields are roughly transverse to the direction of the radio outflow all the way from the core to the southern lobe. It is difficult to interpret the geometry of the inferred B-fields at position CJ1, since the direction of the radio outflow in that region is not clear.

We find that while the southern lobe is co-spatial with the luminous [O~{\sc{iii}}] emission \citep[see also][]{Jarvis19}, the polarized knot J1 exhibits an anti-correlation with [O~{\sc{iii}}] emission (see Figure~\ref{fig5}, right panel), suggesting that the radio emission may be depolarized by the emission-line gas. It is also interesting to note that the [O~{\sc{iii}}] emission extends well beyond the southern lobe towards the south. 

\subsection{J1430+1339}
\label{J1430}
\begin{figure*}
\centerline{
\includegraphics[width=10cm]{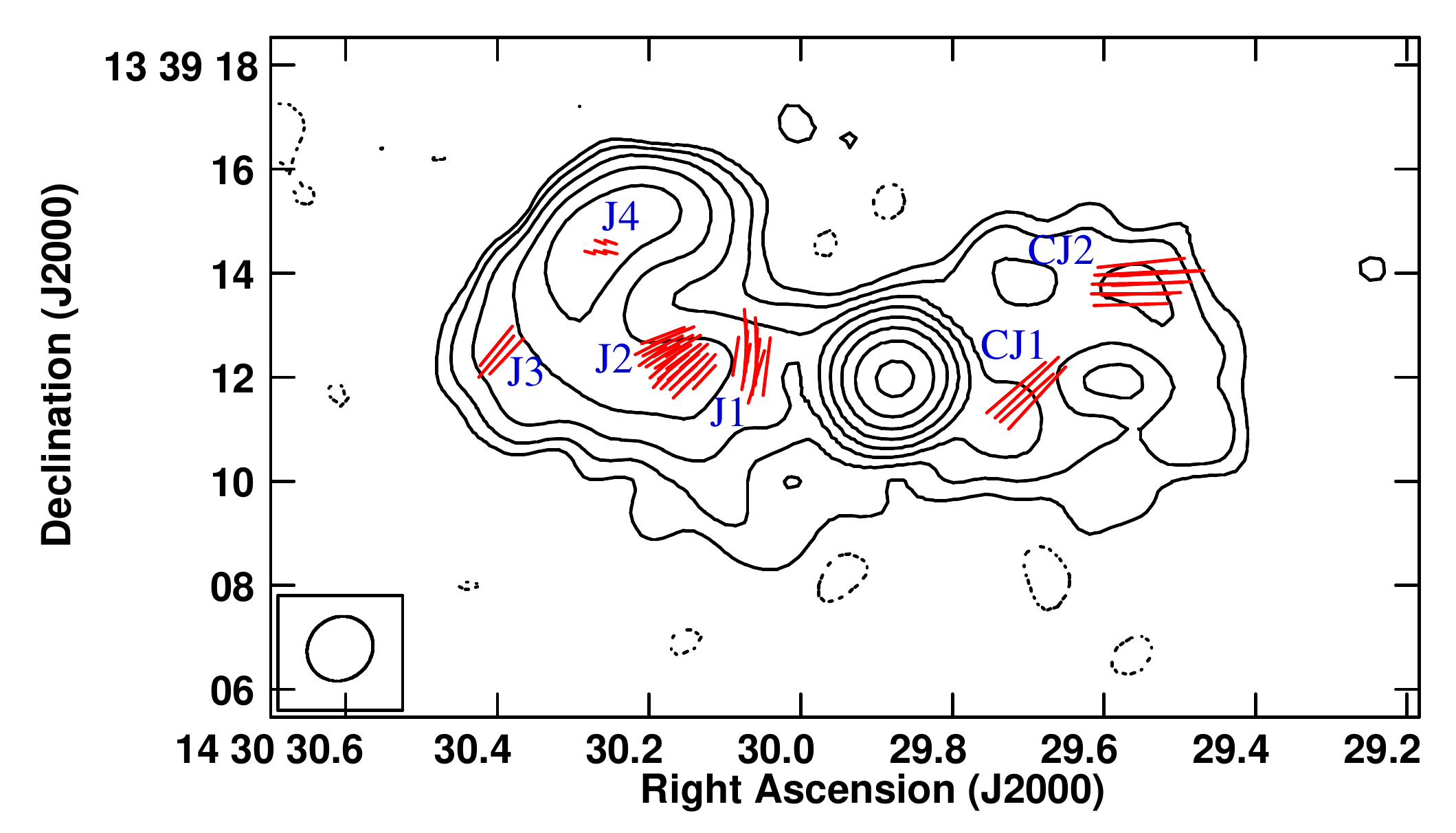}
\includegraphics[width=10.5cm, trim=0 30 0 0]{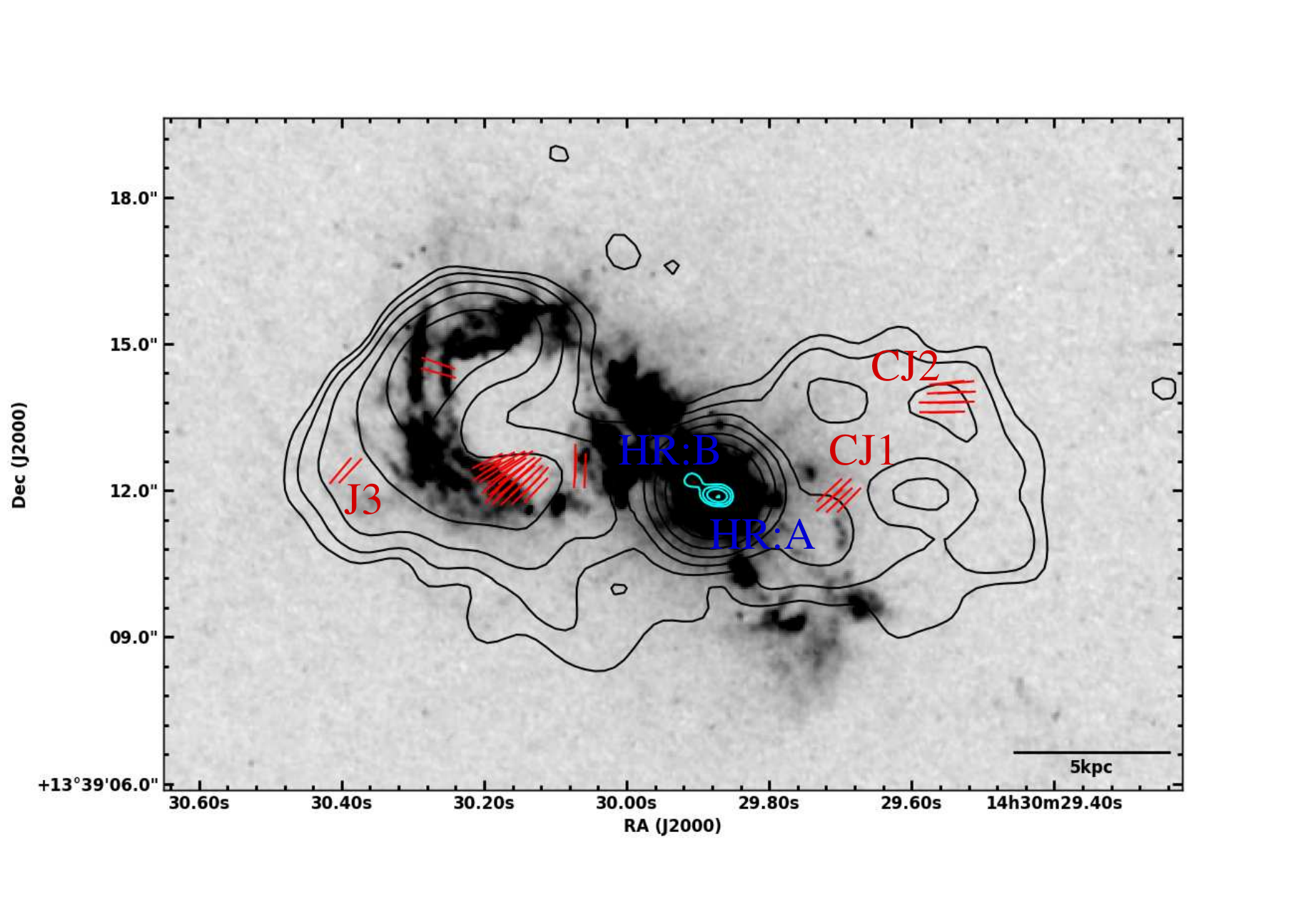}}
\caption{VLA 5 GHz B-array (Beam: 1.31 $\times$ 1.19 arcsec, $-52.02\degr$) total intensity contours in black for J1430+1339 with (left) red superimposed electric polarization vectors, (right) gray-scale HST [O~{\sc{iii}}] image. In the left panel, the length of the vectors are proportional to fractional polarization with 1$\arcsec$ corresponding to 19.2\% fractional polarization. In the right panel, the VLA 6 GHz A-array total intensity contours are in cyan with the polarization vectors proportional to polarised intensity. The peak contour surface brightness is 3.0 mJy beam$^{-1}$ and the levels are 0.019 $\times$ (1, 2, 4, 8, 16, 32, 64, 128, 256, 512) mJy beam$^{-1}$ for the VLA 5 GHz B-array image. The contour levels are 0.023 $\times$ (4, 8, 16, 32, 64) mJy beam$^{-1}$ for the VLA 6 GHz A-array image.}
\label{fig6}
\end{figure*}

This source has been referred to as the ``Teacup' quasar by \citet{Keel12}. We detect double radio lobes/bubbles oriented east-west of the core in this source (black contours in Figure~\ref{fig6}), which have also been detected in the $\sim1$~arcsec image of \citet{Harrison14}. A core-jet structure oriented NE-SW (HR:A and HR:B) at the base of the $\sim1$~arcsec eastern lobe/bubble is detected in the $\sim0.25$~arcsec image of \citet{Jarvis19} (cyan contours in Figure~\ref{fig6}, right panel).

The polarized regions appear as discrete knots (J1, J2, J3, J4, CJ1 and CJ2) along the lobes, with $\chi$ vectors significantly changing in orientation (see Figure~\ref{fig6}). These knots are found near the steep intensity gradients and reveal high fractional polarization: $4-30$\%, with an error typically of $\pm1-10$\%. 

Spatial correlation between radio emission, [O~{\sc{iii}}] emission and X-ray emission along the edges of the eastern lobe of J1430+1339 has already been reported by \citet{Harrison15} and \citet{Lansbury18}. In addition to this, we also find that the polarized knots J1, J2 and J4 are nearly co-spatial with this luminous eastern arc. Essentially, the knot J4 lies in the middle of the two [O~{\sc{iii}}] filaments while the knots J1 and J2 partially overlap with [O~{\sc{iii}}]. The knot J3 exists beyond this arc and is located at a much fainter region of [O~{\sc{iii}}] emission (see Figure~\ref{fig6}, right panel). While on the counter-jet side, the polarized knot CJ1 is also located at a fainter region of [O~{\sc{iii}}] emission whereas CJ2 region exists at a location devoid of [O~{\sc{iii}}] emission. Thus, we find  a clear anti-correlation between the polarized knots J3, CJ1 and CJ2, and [O~{\sc{iii}}] emission, possibly suggesting depolarization of radio emission by emission-line gas.

On the other hand, the spatial coincidence of the radio and [O~{\sc{iii}}] emission along the eastern arc in J1430+1339 suggests significant amount of mixing between the thermal ionized gas and the synchrotron plasma. Therefore, the deficit of polarized emission along the eastern arc can be explained as a result of internal Faraday depolarization of the lobe emission by the co-spatial emission-line gas, although depolarization due to external effects also cannot be ruled out.

\begin{table*}
\begin{center}
\caption{Summary of radio polarization properties}
\label{table2}
\begin{tabular}{lccccc}
\hline
Quasar & Component  & Polarized flux & Fractional & Polarization\\
&  & density & polarization & angle \\
&  & ($\mu$Jy) & (\%) & ($\chi$, $\degr$)\\
\hline
& Core & 27$\pm$6 & 0.5$\pm$0.1  & $-14\pm$7 \\
J0945+1737 & J1 & 1.3$\pm$0.4 & 4$\pm$1  & 33$\pm$8  \\ 
& J2 & 44$\pm$8 & 9$\pm$2 & 27$\pm$6 \\ 
\hline
& Core & 38$\pm$7 & 0.6$\pm$0.1  & $-13\pm$6 \\ 
J1000+1242 & J1 & 10$\pm$3 & 2.0$\pm$0.6  & 28$\pm$9 \\
& J2 & 2.0$\pm$0.6  & 16$\pm$5 & $-51\pm$9  \\
\hline
& Core & 73$\pm$11 & 0.8$\pm$0.1 & 70$\pm$4  \\
J1356+1026 & J1 & 4$\pm$1 & 20$\pm$6  & 45$\pm$9 \\
& CJ1 & 12$\pm$3 & 19$\pm$5  & $-40\pm$8   \\
\hline
& J1 & 5$\pm$2  & 19$\pm$6 &  $-12\pm$9\\
J1430+1339 & J2 & 13$\pm$4 & 14$\pm$4 & $-50\pm$8 \\
& J3 & 1.6$\pm$0.5  & 19$\pm$6  &  $-39\pm$9\\
& J4 & 2.6$\pm$0.8  & 4$\pm$1 & $74\pm$9\\
& CJ1  & 4$\pm$1 & 30$\pm$9 & $-48\pm$9\\
& CJ2  & 7$\pm$2 & 30$\pm$9 &  $-87\pm$8\\
\hline
\end{tabular}
\end{center}
Note: J1010+1413 is excluded from this table since it is unpolarized in current data.
\end{table*}

\begin{table*}
\begin{center}
\caption{Summary of radio total intensity and spectral index properties}
\label{table3}
\begin{tabular}{lccccccc}
\hline
Quasar & Component$^\mathrm{I}$ & Component$^\mathrm{II}$ & $\alpha^\mathrm{I}$ & $\alpha^\mathrm{II}$ & {\it S}$_\mathrm{5GHz}$ & {\it L}$_\mathrm{5GHz}$ & {\it P}$_\mathrm{jet}$ \\
& & & & & (mJy) & (10$^{40}$ erg~s$^{-1}$) & (10$^{44}$ erg~s$^{-1}$) \\
\hline
& Core & LR:A & $-1.04\pm0.01$ & $-0.927\pm0.007$ & 12.5$\pm$0.9 & 2.5 & 4.2 \\
J0945+1737 & NW Lobe & LR:B & $-0.68\pm0.04$ & $-0.74\pm0.03$ & 1.6 & &\\ \hline
& Core & LR:A & $-0.86\pm0.09$ & $-0.52\pm0.05$ & 13$\pm$1 & 3.6 & 5.7 \\ 
J1000+1242 & Jet & LR:B & $-0.95\pm0.05$ & $-0.742\pm0.006$ & 1.0 & &\\
& SE Lobe & LR:C & ... & $-0.5\pm0.1$ & 0.9 & &\\
& NW Lobe & LR:D & ... & ... & 0.8 & & \\
\hline
& Core & LR:A & $-0.83\pm0.03$ & $-0.63\pm0.06$ & 4.0$\pm$0.4 & 2.1 & 3.7 \\
J1010+1413 & N Lobe & LR:B & $-1.3\pm0.1$ & $-0.43\pm0.08$ & 0.6 & & \\
& S Lobe & LR:C & ... & $0.9\pm0.6$ & 0.1 & & \\
\hline
& Core & LR:A & $-1.02\pm0.03$ & $-0.88\pm0.03$ & 19$\pm$2 & 3.5 & 5.6 \\
J1356+1026 & S Lobe & LR:B & $-1.73\pm0.08$ & $-1.1\pm0.8$ & 0.7 & & \\
\hline
& Core & LR:A & $-1.00\pm0.06$ & $-1.02\pm0.01$ & 3.3$\pm$0.3 & 0.3 & 0.7  \\
J1430+1339 & E lobe & LR:B & $-1.2\pm0.2$ & $-0.97\pm0.03$ & 3.1 & & \\
& W Lobe & LR:C & ... & $-0.72\pm0.03$ & 0.9 & & \\
\hline
\end{tabular}
\end{center}
NOTE: Column 1: Source name; Column 2: Radio component as identified in the current work; Column 3: Radio component as identified in \citet{Jarvis19} from their $\sim1\arcsec$ images; Column 4: Mean spectral index estimated in the current work; Column 5: Spectral index reported in \citet{Jarvis19}; Column 6: Total flux density at 5 GHz; Column 7: 5 GHz Luminosity; Column 8: Jet kinetic power estimated using \citet{MerloniHeinz07} relation.
\end{table*}

\begin{table*}
\begin{center}
\caption{Fractional polarization (in \%) and polarization angle (in $\degr$) at the central wavelengths of each sub-band.}
\label{table4}
\begin{tabular}{lccccc}
\hline
Quasar & Component & 0.063 & 0.057 & 0.052 & 0.048 \\
&  & (m) & (m) & (m) & (m) \\
\hline
J0945+1737 & J2 & ($8\pm2$)\%, ($35\pm8)\degr$ & ($9\pm2$)\%, ($22\pm7)\degr$ & ($16\pm5$)\%, ($45\pm9)\degr$ & ($12\pm3$)\%, ($32\pm8)\degr$ \\ 
\hline
\end{tabular}
\end{center}
\end{table*}

\begin{table*}
\begin{center}
\caption{Model parameters for external Faraday depolarization}
\label{table5}
\begin{tabular}{lccccc}
\hline
Quasar & Component & {\it B}$_{||}$ & {\it R} & {\it p}$_{\circ}$ & {\it d} \\
&  & ($\mu$G) & (10$^3$ pc) & (\%) & (10$^{-5}$ pc) \\
\hline
J0945+1737 & J2 & $6.60\pm0.01$ & $2.7\pm1.6$ & $15\pm4$ & $2.8\pm1.7$ \\ 
\hline
\end{tabular}
\end{center}
\end{table*}

\begin{table*}
\begin{center}
\caption{Model parameters for internal Faraday depolarization}
\label{table6}
\begin{tabular}{lccccccccc}
\hline
Quasar & Component & $\alpha$ & {\it p}$_\mathrm{i}$ & {\it B}$_\mathrm{tot}$ & {\it B}$_\mathrm{ord}$ & {\it B}$_\mathrm{ran}$ & {\it L} & {\it n}$_\mathrm{e}$ & {\it M}$_\mathrm{th}$\\
&  &  & (\%) & ($\mu$G) & ($\mu$G) & ($\mu$G) & (pc) & (10$^{-3}$ cm$^{-3}$) & (10$^6$ M$_{\sun}$) \\
\hline
J0945+1737 & J2 & $-0.68\pm0.04$ & $71.6\pm0.5$ & $11.43\pm0.02$ & $5.3\pm0.8$ & $10.1\pm0.4$ & 8860 & $12\pm1$ & $1.01\pm0.08$ \\ 
\hline
\end{tabular}
\end{center}
\end{table*}

\subsection{Modeling depolarization in J0945+1737}
\label{depoln}
We choose J0945+1737 to carry out an analysis of the depolarization mechanisms as its lobe (i.e., region J2) exhibits a strong polarization signal ($>7\%$) in all 4 sub-band polarization images. We created an RM image of J0945+1737 (shown in Figure~\ref{fig2}, bottom panel) using these sub-band polarization images. The average RM in the NW lobe is of the order of 10~rad~m$^{-2}$ with an error of $\sim100$~rad~m$^{-2}$. An RM gradient is observed in a region marginally larger than the beam. However, from this RM value alone, one cannot reliably estimate the cloud sizes or the mixing fraction. Better estimates can result from examining the change in fractional polarization and polarization angle with wavelength. We use the fractional polarization and polarization angle values from sub-band datasets of J0945+1737 for region J2 (see Table~\ref{table4}) in our calculations ahead (see Sections~\ref{ext_depoln} and \ref{int_depoln}). We note that the expected bandwidth depolarization in our $\sim4-6$ GHz band EVLA data is around a factor of 0.5, which is estimated by comparing the ratios of polarized flux density of individual sub-bands (each with a bandwidth of 512 MHz) to that of the full band.

An anti-correlation between radio polarization and [O~{\sc{iii}}] emission has been previously observed in some RL AGN in the literature, that has often been explained as a result of depolarization by the emission-line gas (see Section~\ref{discuss_depoln}). A similar anti-correlation observed in our RQ sources (see Sections~\ref{J0945} to \ref{J1430}) may suggest that the radio emission is being depolarized by a foreground Faraday screen of magneto-ionic media that is clumpy on scales smaller than the observing beam or the emission-line gas that has been entrained in the lobes and mixed with the synchrotron lobe plasma. As suggested in the earlier works on RL AGN (e.g., \citet{O'Sullivan13} for Cen A and \citet{vanBreugel84} for 4C\,26.42), one can estimate the size of the emission-line gas clouds and the amount of ionized gas mixed with the synchrotron lobe plasma in RQ AGN by modeling the external and internal depolarization effects respectively. 

Due to Faraday rotation, the intrinsic polarization angle ($\chi_{\circ}$) differs from the observed polarization angle ($\chi$) by an amount that depends on the square of the wavelength, with the constant of proportionality being the RM, as given below:
\begin{equation}
\mathrm{\chi = \chi_{\circ}~+~RM~\lambda^2},
\end{equation}
where,
\begin{equation}
\mathrm{RM = 812 \int {\it n}_e\,{\it B}_\parallel\,{\it dl}~~~rad~m^{-2}},
\end{equation}
where {\it B}$_\parallel$ is the strength of the line-of-sight component of the ambient magnetic field in milliGauss (mG), {\it n}$_{e}$ is the electron number density of the Faraday-rotating plasma in cm$^{-3}$ and {\it dl} is the path length of the Faraday screen in parsec (pc).
Faraday-rotating particles are usually assumed to be thermal electrons due to the inverse dependence of RM on the square of the particle masses and the effective masses of relativistic electrons being higher than those of the thermal electrons. 

\subsubsection{External depolarization}
\label{ext_depoln}
Faraday-rotating medium external to the synchrotron-emitting plasma acts like a foreground screen that depolarizes the radio emission, for e.g., emission-line gas clouds along the boundaries of the radio lobes. External depolarization also results from instrumental limitations, such as bandwidth and beam depolarization. Bandwidth depolarization occurs due to significant rotation of the polarization angle across an observing bandwidth and this effect is unavoidable in observations since real emission is never fully monochromatic. Beam depolarization occurs due to the presence of B-field and/or electron density fluctuations in the foreground screen on scales less than the telescope beam.

In the case of external depolarization, the fractional polarization at wavelength $\lambda$, i.e. {\it p}($\lambda$) is given as \citep{Burn66}:
\begin{equation}
\mathrm{{\it p}(\lambda) = {\it p}_{\circ}~e^{-2\,(812)^2\,({\it n}_e\,{\it B}_\parallel)^2\,{\it d}\,{\it R}\,\lambda^4}},
\end{equation} 
where {\it p}$_{\circ}$ is the intrinsic fractional polarization, {\it d} (in pc) represents the scale of random fluctuations in the B-field and/or electron density and {\it R} represents the line-of-sight depth through the depolarizing medium. 
This model assumes that these fluctuations occur on scales smaller than the observing beam. This in turn sets an upper limit on the clump/cloud size. Furthermore, it also assumes {\it d} $\ll$ {\it R} and the Faraday dispersion function being represented by a Gaussian.

We assume here an irregular Faraday screen of emission-line gas that is clumpy on scales smaller than the observing beam. In this case, the scale of the electron density fluctuations (i.e., {\it d}) could equivalently represent the scale size of these clumps (such as emission-line gas clouds/filaments) and the line-of-sight depth through the emission-line gas (i.e., {\it R}) could be assumed to be the size of the emission-line region. \citet{Harrison14} provides the radius of the emission line outflow for our sources as the observed semi-major axis of an ellipse fitted to the [O~{\sc{iii}}] region. We use this value for {\it R} in our calculations ahead, considering it to be the size of the emission-line region.

The values of {\it p}($\lambda$) at respective central wavelengths ($\lambda$) of individual sub-bands for the J2 region in J0945+1737 are presented in Table~\ref{table4}. We used equation (4) to fit these data and estimate the best-fitting values of {\it p}$_{\circ}$ and {\it d}. For this, we assumed {\it n}$_\mathrm{e}$ = 100~cm$^{-3}$ \citep[e.g,][]{Liu13,Harrison14}, {\it R} as the radius of the emission line outflow \citep[from Table 4 of][]{Harrison14} and {\it B}$_\parallel$ as the equipartition B-field in the lobes. The equipartition B-field is estimated using the ``minimum'' energy relations from \citet{O'DeaOwen87}, assuming a cylindrical geometry with volume filling factor and proton-to-electron energy ratio equal to 1.0, a spectral index value of $-0.68\pm0.04$ (which is the mean spectral index value of the NW lobe from its in-band spectral index image; see Table~\ref{table3}) and the integrated lobe flux density of 1.6 mJy (see Table~\ref{table3}). The radio spectrum is assumed to extend from 10 MHz to 100 GHz. The value of {\it d} turns out to be $\sim(2.8\pm1.7)\times10^{-5}$ pc. The values of {\it B}$_\parallel$, {\it R}, {\it p}$_{\circ}$ and {\it d} are presented in Table~\ref{table5}. We note that the estimation of {\it d} largely depends on {\it B}$_\parallel$ which we assume here as the equipartition value. However, the use of more constrained value of {\it B}$_\parallel$ will improve our estimation of {\it d}.

\subsubsection{Internal depolarization}
\label{int_depoln}
Internal depolarization occurs when the Faraday-rotating material is mixed with the radio-emitting plasma, for e.g., entrained thermal gas in the radio lobes. Due to the spatial extent of the source, the emission arising at different depths undergo different amounts of Faraday rotation, resulting in a reduction of the fractional polarization of the integrated emission. Internal depolarization accounts for both ordered B-field ({\it B}$_\mathrm{ord}$) and random B-field ({\it B}$_\mathrm{ran}$) components. We consider a simple case where the source is assumed to be a slab with a linear depth {\it L} along the line of sight, {\it d} represents the scale of fluctuations of the random B-field component and a Gaussian distribution of Faraday depths is assumed \citep{Burn66, Sokoloff98}. In this case, the complex polarization {\it P}($\lambda$) is given as:
\begin{equation}
\mathrm{{\it P}(\lambda) = {\it p}_{\circ}~(\frac{1-e^{-{\it S}}}{{\it S}})},
\end{equation}
where 
\begin{equation}
\mathrm{{\it S} = 2\,(812\,{\it n}_e\,{\it B}_{ran})^2\,{\it d}\,{\it L}\,\lambda^{4} - 2i\,(812\,{\it n}_e\,{\it B}_{ord})\,{\it L}\,\lambda^{2}}
\end{equation}

The intrinsic fractional polarization for synchrotron emission with ordered B-fields and optically thin emission is reduced from its maximum value ({\it p}$_\mathrm{i}$) according to the relation: $\mathrm{{\it p}_{\circ} = {\it p}_{i}({{\it B}^2_{ord}}/{{\it B}^2_{tot}})}$, where 
$\mathrm{{\it B}^2_{tot}={\it B}^2_{ord}+{\it B}^2_{ran}}$ and $\mathrm{{\it p}_{i} = (3-3\alpha})/({5-3\alpha})$, $\alpha$ being the spectral index \citep{Burn66}. We make use of these relations to estimate {\it B}$_\mathrm{ord}$ and {\it B}$_\mathrm{ran}$ for the NW lobe of J0945+1737. For this, we first estimate {\it p}$_\mathrm{i}$ by taking $\alpha$ as $-0.68\pm0.04$, and then estimate {\it B}$_\mathrm{ord}$ and {\it B}$_\mathrm{ran}$ by taking {\it B}$_\mathrm{tot}$ as \,$\sqrt[]{3}${\it B}$_\parallel$ (from geometry) and {\it p}$_{\circ}$ as obtained in Section~\ref{ext_depoln}.

We use the same value of {\it d} as obtained in Section~\ref{ext_depoln} and {\it L} as the length of the lobe. Solving equation (5) using {\it p}($\lambda$), $\chi$ (such that {\it P}($\lambda$) = {\it p}($\lambda$)e$^{2i\chi}$), and $\lambda$ from Table~\ref{table4} for individual sub-bands, with the Python package LMFIT \citep{Newville14}, we obtain the value of {\it n}$_\mathrm{e}$ as $\sim(12\pm1)\times$10$^{-3}$ cm$^{-3}$. 

The total mass of the thermal gas within the lobes $M_\mathrm{th}$ is given as:
\begin{equation}
M_\mathrm{th}\sim{n}_\mathrm{e}\,{m}_\mathrm{H}\,{f}_\mathrm{V}\,{V}
\end{equation}
where, $m_\mathrm{H}$ is the mass of the ionized hydrogen ($\sim1.67\times10^{-24}$ g), {\it V} is the volume of the lobe and {\it f}$_\mathrm{V}$ is the volume filling factor. We derive {\it V} $\sim1\times10^{67}$ cm$^{3}$ for the NW lobe of J0945+1737 assuming a cylindrical geometry. We use {\it n}$_\mathrm{e} \sim(12\pm1)\times$10$^{-3}$ cm$^{-3}$ as estimated above and {\it f}$_\mathrm{V}\approx$ 10$^{-2}$ \citep{Osterbrock91}. This yields {\it M}$_\mathrm{th}$ $\sim(1.01\pm0.08)\times10^{6}$~M$_{\sun}$ (more in Section~{\ref{discuss_depoln}}). The values of $\alpha$, {\it p}$_\mathrm{i}$, {\it B}$_\mathrm{tot}$, {\it B}$_\mathrm{ord}$, {\it B}$_\mathrm{ran}$, {\it L}, {\it n}$_\mathrm{e}$ and {\it M}$_\mathrm{th}$ are provided in Table~\ref{table6}.

\section{Discussion}
\label{discuss}
\subsection{Origin of radio emission}
\label{discuss_origin}
Polarization is detected in 4 of 5 sources in the current data. Polarization is not detected in J1010+1413. The cores exhibit a fractional polarization of $0.5-1$\%, which is typical even for RL AGN, whereas the lobes reveal high fractional polarization ($10-30\%$).
The inferred B-fields in the core of J0945+1737, roughly aligned with the local direction of the $\sim1\arcsec$ jet, may be consistent with a jet threaded by poloidal B-fields \citep[e.g.][]{MehdipourConstantini19, Silpa21a, Silpa21b}. The complexity associated with the spatial overlap of HR:C with the polarized knot J2 in J0945+1737 may explain why the inferred B-fields at position J2 are neither parallel nor perpendicular to the local jet direction. These signatures are unlikely to appear in case of a non-jet origin for radio emission. The transverse B-fields in the core and knot J2 in J1000+1242 are suggestive of internal shocks that order and amplify B-fields by compression. Ordered B-fields and high polarization in J1000+1242 does not favour a radiatively-driven wind origin for the radio emission, thus making the jet interpretation for this source stronger and also consistent with the result of \citet{Jarvis19}. The presence of $\sim0.25\arcsec$ core-jet and hotspot features also favour a jet origin in these sources.

The roughly transverse B-fields all the way from the core to the southern lobe in J1356+1026 could either represent a series of transverse shocks that order and compress B-fields like in BL Lac jets or a toroidal component of a large-scale helical B-field associated with a jet \citep{Gabuzda94, Lister98, Pushkarev17}. Conversely, the inferred B-field geometry could as well suggest an AGN ``wind'', threading toroidal B-fields \citep[e.g,][]{MehdipourConstantini19, Silpa21a, Silpa21b}. This `wind' component could either be an accretion disk wind \citep{MehdipourConstantini19, Miller12} or the outer layers of a broadened jet \citep[resembling a jet sheath;][]{Mukherjee18} or a combination of both. The wind origin is further supported by the absence of a $\sim0.25\arcsec$ jet in this source. Thus, it is difficult to fully discern the origin of radio emission (jet or wind) in this source with the current data.

J1430+1339 resembles the radio galaxy 4C 26.42 in its polarization properties \citep{vanBreugel84}. Knotty polarization structures in this source suggest ordered B-fields on smaller length scales. This is indicative of a turbulent jet with small-scale Kelvin-Helmholtz (KH) instabilities \citep{Mukherjee20}. The growth of KH instabilities could either suggest that the B-fields at the jet base are not strictly toroidal or that the toroidal component of a large-scale helical B-field associated with the jet is not strong enough to suppress these instabilities. This may be consistent with the inferred B-fields aligned with the local direction of the $\sim1\arcsec$ radio outflow at position J1. The presence of KH instabilities also facilitates mixing of the jet plasma with the ambient medium, resulting in the deceleration and decollimation of the jet. This may be the case in J1430+1339, where the $\sim0.25\arcsec$ jet may have lost collimation and broken out into $\sim1\arcsec$ lobes/bubbles. The jet kinetic power of $\sim$ 7$\times$10$^{43}$~erg~s$^{-1}$ as estimated using equation (1) suggests a low power jet in this source, consistent with \citet{Mukherjee20} where the authors suggest that the low power jets are more susceptible to instabilities.

The knotty nature of polarization structures with high degrees of polarization in J1430+1339 could be explained as a beam depolarization effect. This can be tested with polarization observations at lower resolution, wherein we expect several such knots with different polarization angles within the telescope beam to sum up and give lower degrees of polarization and continuous polarized regions \citep[e.g.][]{vanBreugel84}.

The morphological and spectral properties of J1430+1339 such as the bubble-like lobes with ultra-steep spectrum, mis-alignment between the $\sim0.25\arcsec$ jet and the $\sim1\arcsec$ lobes, and radio emission on different spatial scales, are similar to those observed in typical Seyfert galaxies powered by an episodic precessing jet \citep[for e.g. Mrk 6;][]{Kharb06}. Alternatively, these morphological features can also be produced in a model prescribed by \citet{Mukherjee18} using numerical simulations. Their model demonstrates that the strong interaction of an inclined relativistic jet with the dense ISM produces a sub-relativistic outflow reminiscent of an AGN ``wind'' \citep{RupkeVeilleux11} or a nuclear starburst \citep{Veilleux05}, along the minor axis of the host galaxy following a path of least resistance. Their model predicts that such a ``wind'' creates a spherical bubble that clears the ambient medium as it rises and the jet eventually proceeds roughly along its axis of launch to some extent. On large scales the structure might resemble a wind, but it is in fact jet-driven as one would find a jet closer to the nucleus, similar to the case in J1430+1339.

Overall, we find that the origin of radio emission in J0945+1737, J1000+1242, J1010+1413 and J1430+1339 is likely to be jets based on their $\sim0.25\arcsec$ jet-like features, spectral properties (see Appendix~\ref{appendixA}) and polarization properties (except for J1010+1413). On the other hand, the origin of radio emission cannot be clearly interpreted (jet or wind) in case of J1356+1026 from the current data.

\subsection{Implications of the depolarization effects}
\label{discuss_depoln}
All sources in our sample are known to exhibit a spatial correlation between radio emission and [O~{\sc{iii}}] emission \citep[e.g.,][]{Jarvis19}. The current work suggests an anti-correlation between various polarized knots and [O~{\sc{iii}}] emission in these sources. Essentially, we find polarization mostly in the jet/lobe regions that are deficit of [O~{\sc{iii}}] emission and its absence in regions dominated by [O~{\sc{iii}}] emission. This is similar to that seen in some RL AGN in the literature. 

A preferential occurrence of bright optical line emission along the radio source boundary and its anti-correlation with the polarized radio emission has been commonly found in radio galaxies. This is often explained by the depolarization of the radio emission by an irregular Faraday screen of clumpy magneto-ionic thermal gas surrounding the source, with typical clump sizes much smaller than the observing beam. For example, \citet{Burn66}, \citet{Heckman82} for 3C 305, \citet{vanBreugelHeckman82}, \citet{vanBreugel84} for 4C 26.42, \citet{vanbreugel85a} for 3C 277.3, \citet{vanbreugel85b} for Minkowski's object, \citet{Heckman84} for 3C 171, \citet{Clark97} for PKS 2250-41, \citet{Hardcastle03} for 3C 171. On the other hand, strong depolarization most likely due to thermal gas mixed with the relativistic plasma in the lobes of the radio galaxy Cen A has been reported by \citet{O'Sullivan13}.

We estimate the size of the emission-line gas clouds in the J2 region of J0945+1737 to be $\sim(2.8\pm1.7)\times10^{-5}$ pc by modeling external depolarization of the lobe emission. The scale size of the fluctuations in the lobes of radio galaxies have been reported to lie in the range of $10^{-5}-300$ pc (the upper limit being set by half the beam size at 6 cm; see \citet{vanBreugel84} for 4C\,26.42, \citet{vanbreugel85a} for 3C\,277.3). The upper limit on {\it d} for J0945+1737 turns out to be $\sim$1200 pc with the current data, implying a scale size range of 10$^{-5}$ pc $\leq$ {\it d} $\leq$ 1300 pc for the emission-line clouds in this source.
 
We also estimate the amount of thermal material mixed with the synchrotron plasma in the lobe of J0945+1737, which turns out to be $\sim(1.01\pm0.08)\times10^{6}$~M$_{\sun}$, by modeling internal depolarization of the lobe emission. Much higher estimates of mixed thermal gas masses ($\sim10^9-10^{10}~{\mathrm M_{\sun}}$) have been found in the lobes of radio galaxies using X-ray and CO observations. For example, \citet{Simionescu09} for Hydra A, \citet{Kirkpatrick09} for the brightest cluster galaxy in Abell 1664,
\citet{Salome11} for NGC 1275, \citet{Werner11} for the core of galaxy cluster S{\'e}rsic 159-03, \citet{Russell17} for the central galaxy of the Phoenix cluster and \citet{Anderson18} for Fornax A. See also the review by \citet{McNamaraNulsen12}. \citet{O'Sullivan13} find thermal gas mass of $\sim10^{10}~\mathrm M_{\sun}$ in the lobes of the radio galaxy Cen A using depolarization analysis with $\sim1-2$ GHz data. 

Similar studies have not been widely carried out for RQ AGN in the literature. Nevertheless, some recent studies have found ionized gas masses of the order of $10^5-10^7$~M$_{\sun}$ on few kpc-scales in Seyfert 2 galaxies using optical spectroscopic and imaging data \citep[e.g.,][]{Fraquelli03, Revalski21}. \citet{Yoshida02} find ionized gas mass of $\sim10^5~\mathrm M_{\sun}$ and clouds/filaments size of the order of 100 pc in a $\sim$35 kpc-scale emission line region around the Seyfert type 2 galaxy NGC 4388, using Subaru observations. We note that the ionized gas mass derived for the type 2 RQ quasar J0945+1737 in the current work broadly agrees with the literature.

It is difficult to explain the S-shaped symmetry between radio and [O~{\sc{iii}}] emission as revealed in sources like J1000+1242 and J1010+1413, using a simple photo-ionization model. On the other hand, a shock-ionization model seems more likely, where the local emission-line gas is accelerated, compressed, heated and ionized due to shocks driven by relativistic jets or radiatively-driven AGN winds. However, our current findings favour jet-driven shocks as primary contenders in both these sources, which could be tested with IFU data in the future.

\subsection{Caveats}
\label{caveats}
We find that while a clear anti-correlation is found between the polarized emission and [O~{\sc{iii}}] emission in certain regions of our sources (8/12 knots), there are also regions where a correlation is found (4/12 knots). Nevertheless, the polarized knots in these sources that show an anti-correlation with [O~{\sc{iii}}] emission outnumber those that show a correlation. A correlation could be possibly explained as a result of larger clump sizes of the depolarizing media as compared to the observing beam or insufficient mixing with the synchrotron plasma.

It is also difficult to unambiguously rule out other possible reasons for the anti-correlation observed in our sources, such as lack of sensitivity and beam depolarization effects. Based on the redshifts of our sources, we find that the spatial extents of the closest source, (i.e., J1430+1339, z=0.085), will be sampled two times more than that of the farthest source (i.e., J1010+1413, z=0.199) by the current observing beam of $\sim$1 arcsec. Moreover, from the polarization image of J1430+1339, we see that the polarized emission is arising from compact regions with ordered B-fields. We also estimated the typical drop in the fractional polarization due to beam depolarization effects in the NW lobe of J0945+1737. The relative error in the polarized intensity turns out to be $\sim$40\% when the beam is doubled in size. This suggests that if the beam depolarized signal falls below the sensitivity limit of our experiment, then polarization cannot be detected. This is likely to be the case in J1010+1413, where the limit on the fractional polarization of the brightest feature, i.e. the core, is $\sim$1\%.

We note that disentangling internal and external depolarisation requires deeper multi-frequency and multi-resolution observations, hence, the current work involving $\sim4-6$ GHz band data has limited implications. However, this study is aimed at providing a first-order analysis of the depolarization mechanisms at play in this source and lower limits on the derived parameters like {\it d} and {\it n}$_\mathrm{e}$. We emphasize that the system is more complex than the models used here, and RM synthesis \citep{BrentjensdeBruyn05} must be used instead.

\subsection{Presence of double nuclei}
\label{double_nuclei}
We note that three sources in our sample: J1000+1242, J1010+1413 and J1356+1026, show 2 or more optical nuclei in their HST WFC3 images. We also find that only one of the optical nuclei has an associated radio source in all 3 sources. 

\begin{figure*}
\centerline{
\includegraphics[width=10cm]{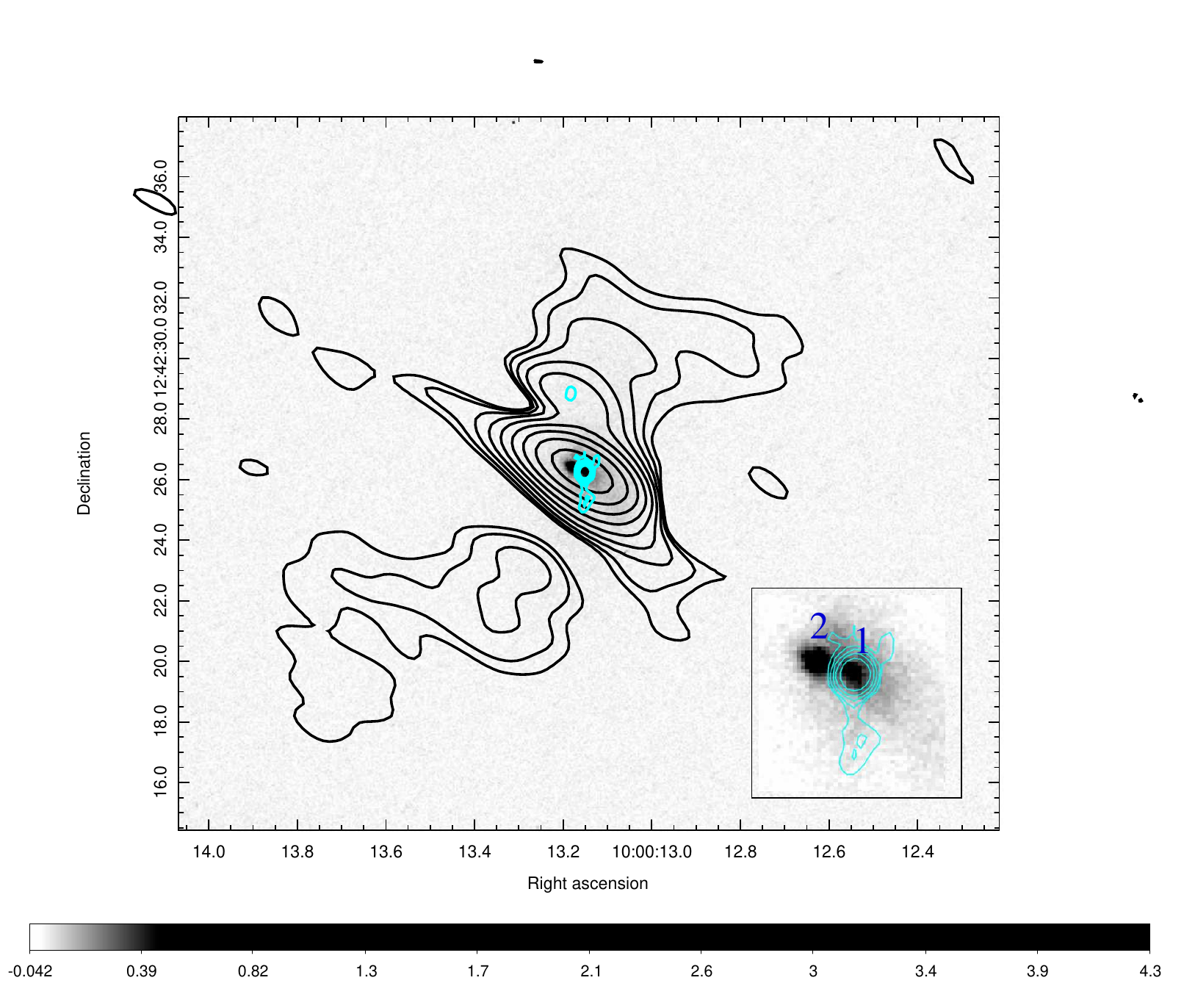}
\includegraphics[width=10.3cm]{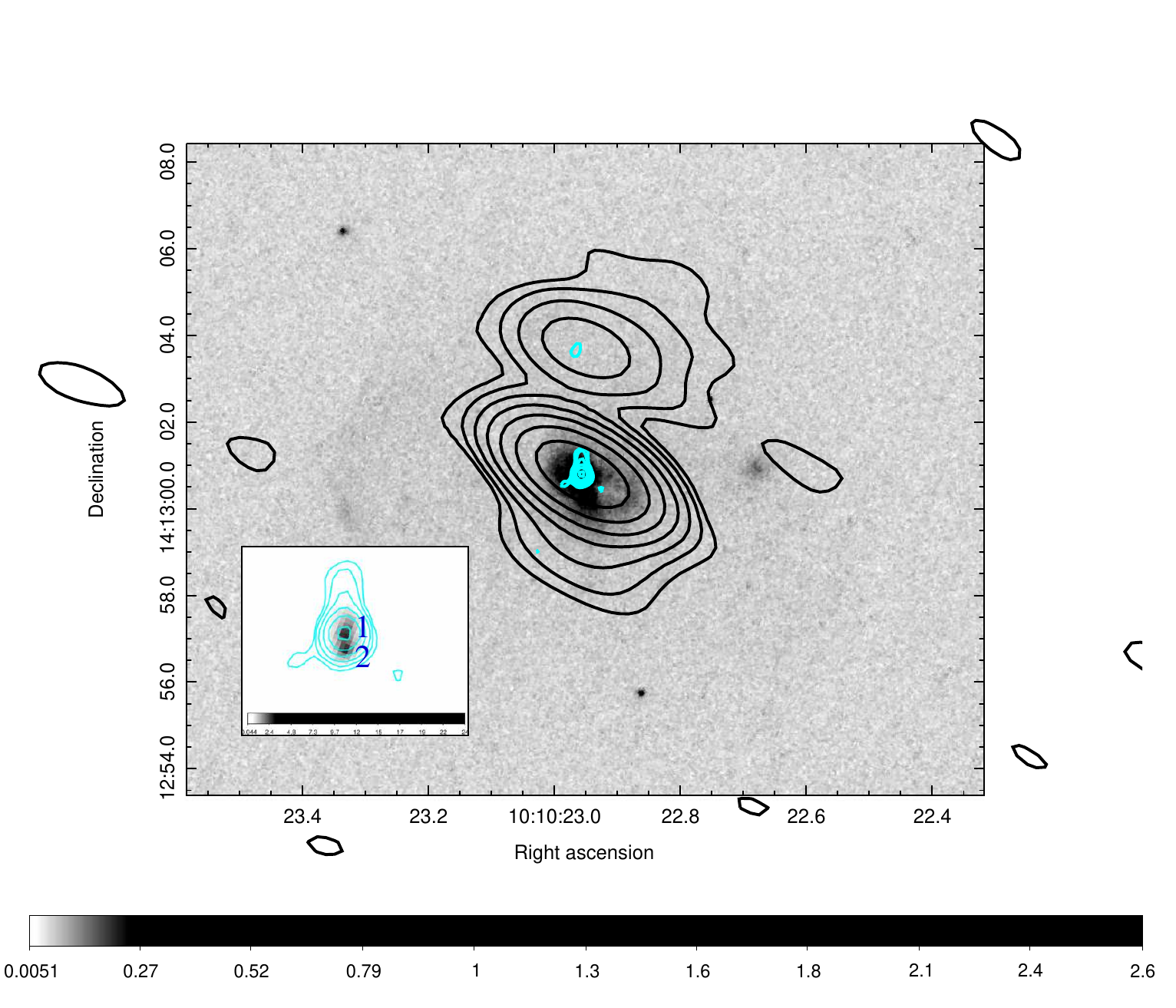}}
\centerline{
\includegraphics[width=10cm]{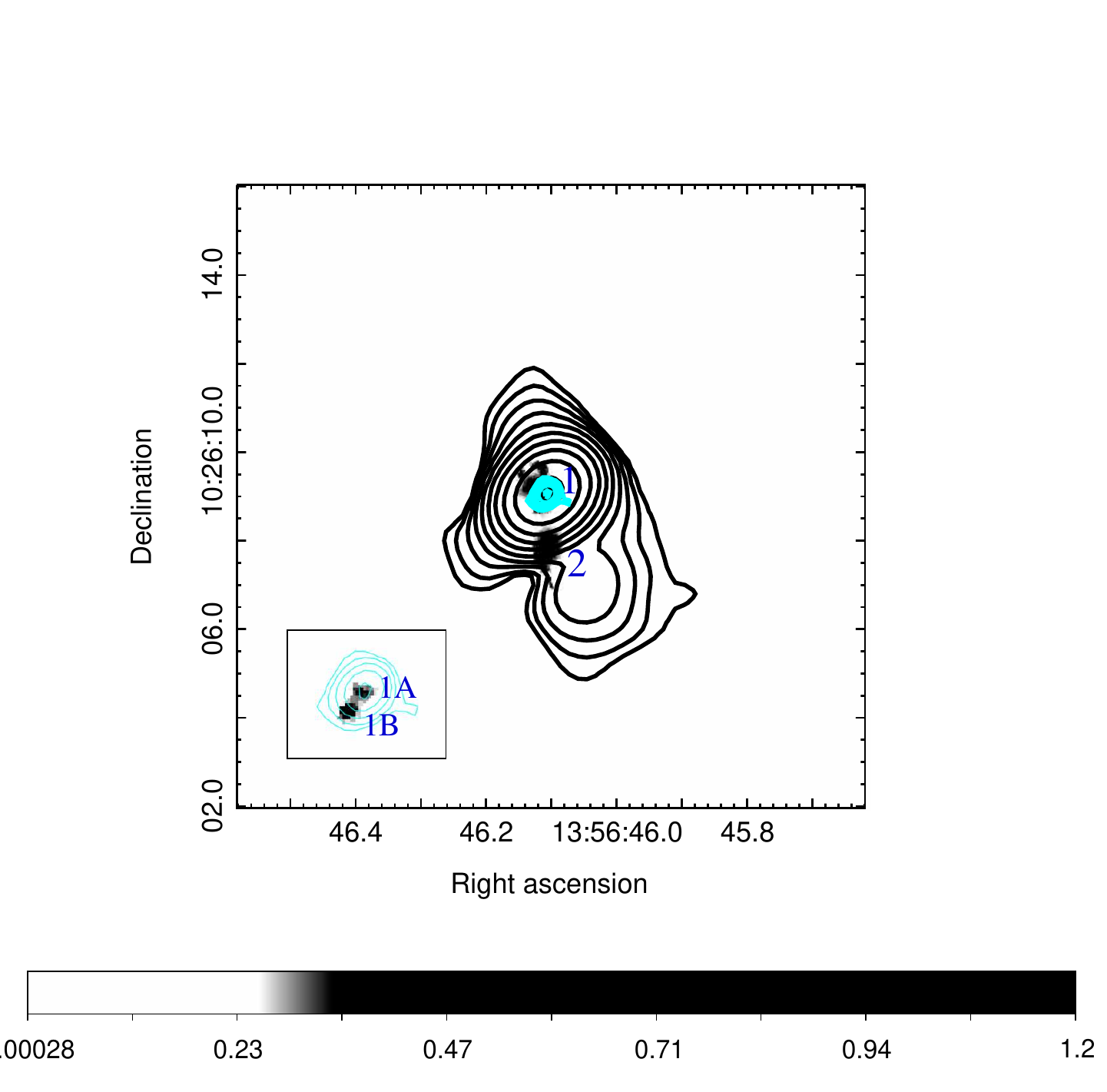}}
\caption{The main panel shows HST WFC3 image in gray-scale superimposed with VLA 5 GHz B-array total intensity contours in black and VLA 6 GHz A-array contours in cyan for Top left: J1000+1242; Top right: J1010+1413; Bottom: J1356+1026. The inset in each panel shows a zoom-in of the core region, indicating the presence of dual optical nuclei. The optical nuclei are marked in blue. The peak contour surface brightness is {\it x} mJy beam$^{-1}$ and the contour levels are {\it y} $\times$ (1, 2, 4, 8, 16, 32, 64, 128, 256, 512) mJy beam$^{-1}$ for VLA 5 GHz B-array image, where ({\it x} ; {\it y}) for J1000+1242, J1010+1413 and J1356+1026 are (12;0.0265), (3.8;0.031) and (18;0.030) respectively. The contour levels are {\it y}$^*$ $\times$ (8, 16, 32, 64, 128) mJy beam$^{-1}$ for VLA 6 GHz A-array image, where {\it y}$^*$ for J1000+1242 and J1356+1026 are 0.023 and 0.1 respectively. The contour levels are 0.016 $\times$ (4, 8, 16, 32, 64, 128) mJy beam$^{-1}$ for VLA 6 GHz A-array image of J1010+1413. The color scale extends from $-0.042$ to $4.3~\times$ ($5.20\times10^{-7}) = -0.022$ to 2.2 $\mu$Jy for J1000+1242. For J1010+1413, the color scale extends from $0.0051$ to $2.6~\times$ ($5.87\times10^{-7}) = 3$ nJy to 1.5 $\mu$Jy in the main panel while it extends from $-0.044$ to $24~\times$ ($5.87\times10^{-7}) = -0.026$ to 14 $\mu$Jy in the inset. The color scale extends from $0.00028$ to $1.2~\times$ ($5.24\times10^{-7}) = 0.1$ nJy to 0.6 $\mu$Jy for J1356+1026. In these images, we have used the HEASARC Fv software (https://heasarc.gsfc.nasa.gov/docs/software/ftools/fv/) to arbitrarily make the closer optical nucleus coincident with the radio core in our images.}
\label{fig7}
\end{figure*}

We find two optical nuclei in J1000+1242 with a projected separation of $\sim$1.1 kpc (marked as `1' and `2' in Figure~\ref{fig7}, top left panel), only one of which has an associated radio source (i.e. `1'). Also, neither the $\sim0.25\arcsec$ VLA image nor the eMERLIN image of this quasar in \citet{Jarvis19} (also see their online supplementary material) shows the presence of a secondary radio source.

\citet{Goulding19} identify a SMBH pair in J1010+1413 with a projected separation of $\sim$430 pc using HST/WFC3 images, though they also suggest a possibility of the quasar scattered light origin. We also identify a pair of optical nuclei in the HST WFC3 image of this quasar with a projected separation of $\sim$0.4 kpc (marked as `1' and `2' in Figure~\ref{fig7}, top right panel) and a radio source associated with one of the nuclei (i.e `1').

A merging pair of AGNs separated by $\sim$2.5 kpc has been identified in J1356+1026 from optical \citep{Liu10,Greene11,Fu11}, near-infrared \citep{Shen11} and X-ray \citep{Comerford15,Foord20} observations. We also find a pair of optical nuclei in the HST WFC3 image of this source with a projected separation of $\sim$2.6 kpc (marked as `1' and `2' in Figure~\ref{fig7}, bottom panel) and the $\sim0.25\arcsec$ radio core being associated with only one of them (i.e. `1'). Interestingly, we also find that the optical nuclei `1' is actually composed of a double nuclei with a projected separation of $\sim$0.4 kpc (marked as `1A' and `1B').

We note that all sources in our sample show evidence for ongoing galactic mergers. Specifically, J1000+1242, J1010+1413 and J1356+1026 show presence of double nuclei in their optical images. J0945+1737 and J1430+1339 are also known to have undergone mergers (for e.g, \citet{VillarMartin21} for J0945+1737 and \citet{Keel15,Harrison15} for J1430+1339). \citet{VillarMartin21} also show that in their full sample of 13 z$<$0.2 type 2 quasars (which includes our sources except J1010+1413), all but one (which is not in our sample) show evidence for mergers and these sources have ionized gas spread over large spatial extents. The mergers may have caused gas and dust to be stripped from the galactic companions making this medium available for the jets/winds to interact with. This increased jet-medium interaction may have led to preferentially higher [O~{\sc{iii}}] luminosities in these sources, which would then have an influence on the sample selection. Ongoing galactic mergers may also influence our interpretation of AGN feedback \citep[e.g.,][]{Jarvis19,VillarMartin21}, in these sources.

We also note that while the distorted lobes in J1000+1242 and the curved lobe in J1356+1026 could be explained as a result of jet precession (both being dual AGN candidates), they could also arise from jet-medium interactions. Similar is the case for J0945+1737, owing to the presence of its curved $\sim1\arcsec$ jet, and the mis-alignment between the $\sim0.25\arcsec$ nuclear features (i.e. HR:A and HR:B) and the polarized knot J2. However, we do not find any signature of a double nuclei in its optical image as an evidence for jet precession. 

\section{Conclusions}
\label{conclusions}
We summarize below the results from our VLA 5 GHz radio polarization study in conjunction with [O~{\sc{iii}}] emission-line study of 5 type 2 radio-quiet quasars. 
\begin{enumerate}
\item
The morphological, spectral and polarization properties suggest a jet origin for radio emission in J0945+1737, J1000+1242, J1010+1413 and J1430+1339 whereas the current data cannot fully discern the origin of radio emission (jet or wind) in J1356+1026.

\item 
4 out of 5 sources in our sample show polarization in the current data. We find an anti-correlation between various polarized knots and [O~{\sc{iii}}] emission in these sources. This is similar to that observed in some radio-loud AGN in the literature. This suggests that the radio emission is likely to be depolarized by an irregular Faraday screen of emission-line gas that is clumpy on scales smaller than the observing beam or by the entrained emission-line gas in the lobes of these sources that has mixed with the synchrotron lobe plasma.

\item 
As J0945+1737 exhibits the most prominent polarization signature in its lobe ($>7\%$ in 4 individual sub-band images), we chose this source to model the external and internal depolarization of lobe emission. Modeling external depolarization yields the size of the emission-line gas clouds in the J2 region of the NW lobe to be $d\sim(2.8\pm1.7)\times10^{-5}$ pc. Setting an upper limit on {\it d} as half the beam size yields a scale size range of 10$^{-5}$ pc $\leq$ {\it d} $\leq$ 1300 pc for the emission-line gas clouds, which broadly spans the range of values derived for radio-loud AGN in the literature. 

By modeling the internal depolarization, the total mass of the thermal gas mixed with the synchrotron plasma in its NW lobe turns out to be $\sim(1.01\pm0.08)\times10^{6}$~M$_{\sun}$. Similar orders of ionized gas masses ($\sim10^5-10^7~{\mathrm M_{\sun}}$) have been derived for a few radio-quiet AGN in the literature using optical spectroscopic and imaging data. However, much higher estimates of mixed thermal gas masses ($\sim10^9-10^{10}~{\mathrm M_{\sun}}$) have been found in the lobes of radio galaxies using X-ray and CO observations as well as radio polarization data.

Our results show that as the radio outflow (jet/lobe/wind) traverses the ambient medium, it entrains gas which mixes with the synchrotron plasma and depolarizes the radio emission. Therefore, the anti-correlation observed between polarization and ionized gas in our sources could be interpreted as an effect of the radio-gas interaction (and a possible source of AGN feedback) in these sources. The current work demonstrates that the interplay of jets/winds and emission-line gas is most likely responsible for the nature of radio outflows in radio-quiet AGN. Furthermore, we suggest that the type 2 quasars show certain similarities to the large-scale radio-powerful AGN.
\end{enumerate}

\section*{Acknowledgements}
We thank the referee for their insightful suggestions. CMH acknowledges support from United Kingdom Research and Innovation (grant reference: MR/V022830/1). SS would like to thank Shane O'Sullivan for providing inputs and scripts for looking at the depolarization relations as used in their 2013 paper. We acknowledge the support of the Department of Atomic Energy, Government of India, under the project 12-R\&D-TFR-5.02-0700. The National Radio Astronomy Observatory is a facility of the National Science Foundation operated under cooperative agreement by Associated Universities, Inc.

\section*{Data Availability}
The data underlying this article will be shared on reasonable request to the corresponding author. The VLA data underlying this article can be obtained from the NRAO Science Data Archive (https://archive.nrao.edu/archive/advquery.jsp) using the proposal ID 20A-176. The HST [O~{\sc{iii}}] images underlying this article can be obtained from the MAST HST Archive (https://archive.stsci.edu/hst/search.php) using the proposal ids: JCLN07030 for J0945+1737, ID6T01010 for J1000+1242, ID6T02010 for J1010+1413, IBWE02020 for J1356+1026, JBQVA2020 for J1430+1339. The HST optical images underlying this article can be obtained from the MAST HST Archive (https://archive.stsci.edu/hst/search.php) using the proposal ids: ID6T01020 for J1000+1242, ID6T02020 for J1010+1413 and ICRT01010 for J1356+1026.



\bibliographystyle{mnras}
\bibliography{mnras} 




\appendix
\section{IN-BAND SPECTRAL INDEX IMAGES}
\label{appendixA}
While imaging with the MT-MFS algorithm of the $\tt{TCLEAN}$ task in $\tt{CASA}$, we used two Taylor terms to model the frequency dependence of the sky emission by setting the parameter $\tt{nterms}$=2. This produced in-band spectral index images and spectral index noise images for individual sources. We blanked the pixels with spectral index errors greater than 0.09 for J0945+1737, 0.1 for J1000+1242, 0.18 for J1010+1413, 0.15 for J1356+1026 and 0.4 for J1430+1339, which span a range between 10 and 20\%.

\begin{figure*}
\centerline{
\includegraphics[width=9.5cm, trim=0 200 0 200]{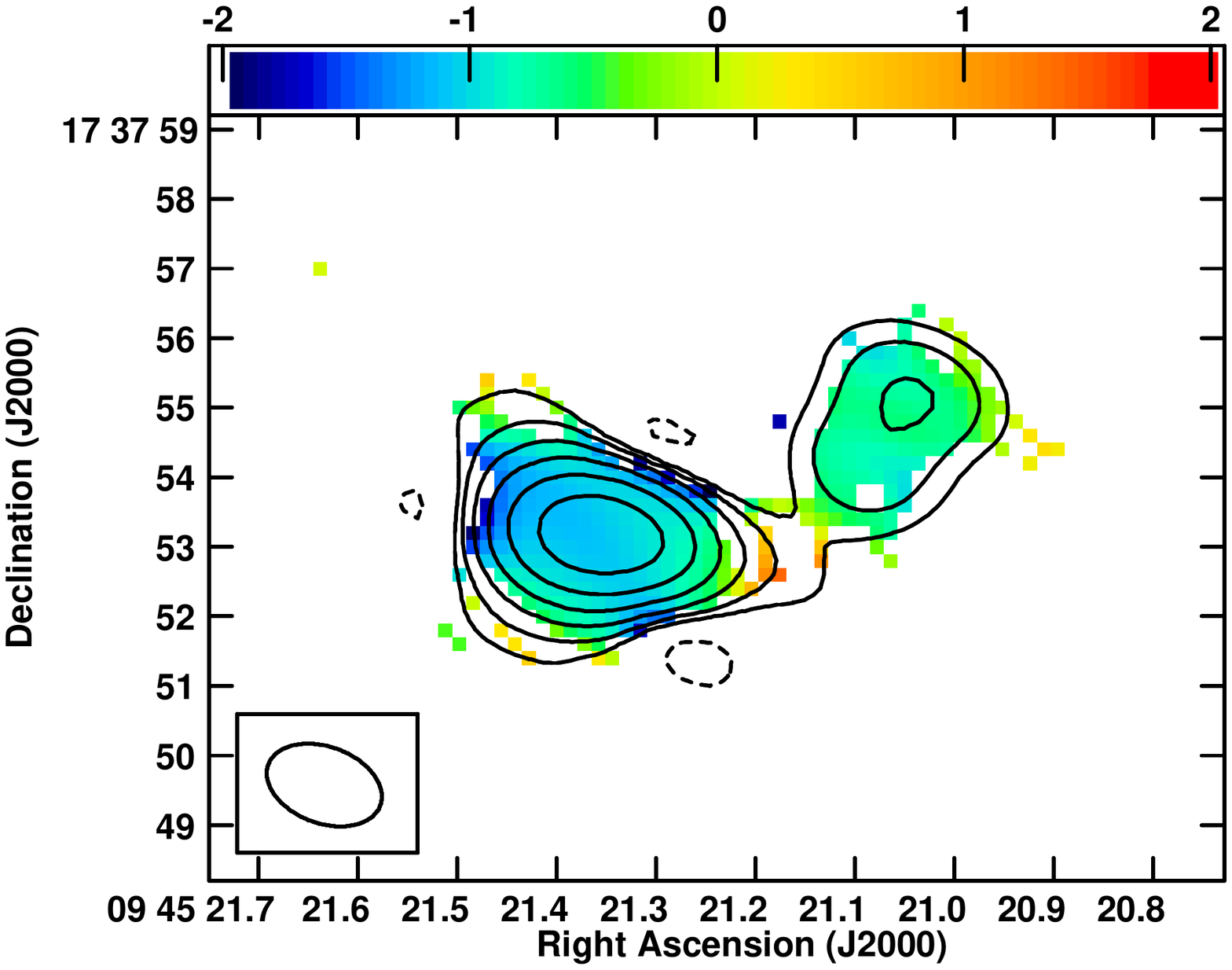}
\includegraphics[width=9.3cm, trim=0 200 0 200]{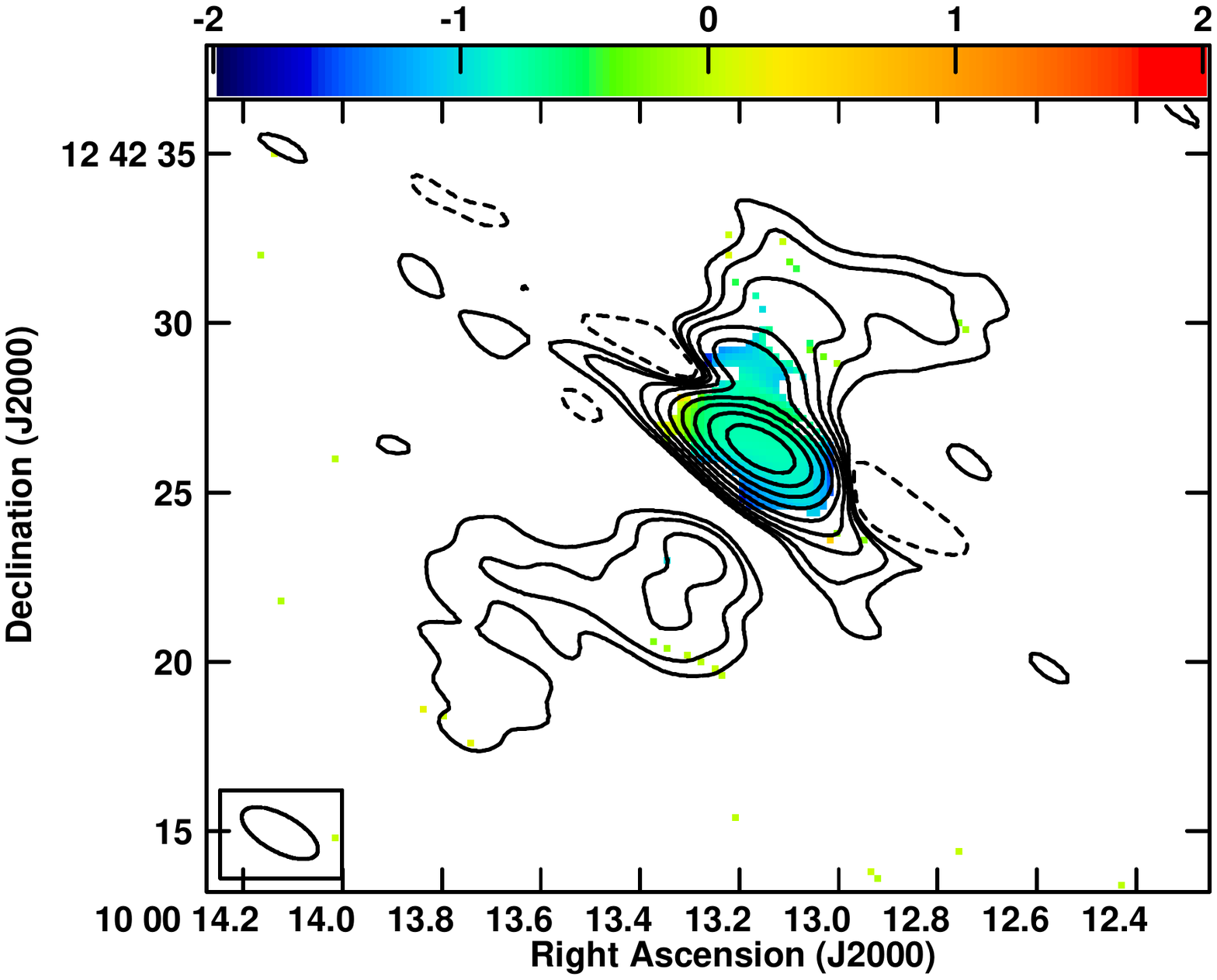}}
\centerline{
\includegraphics[width=8.3cm, trim=0 200 0 200]{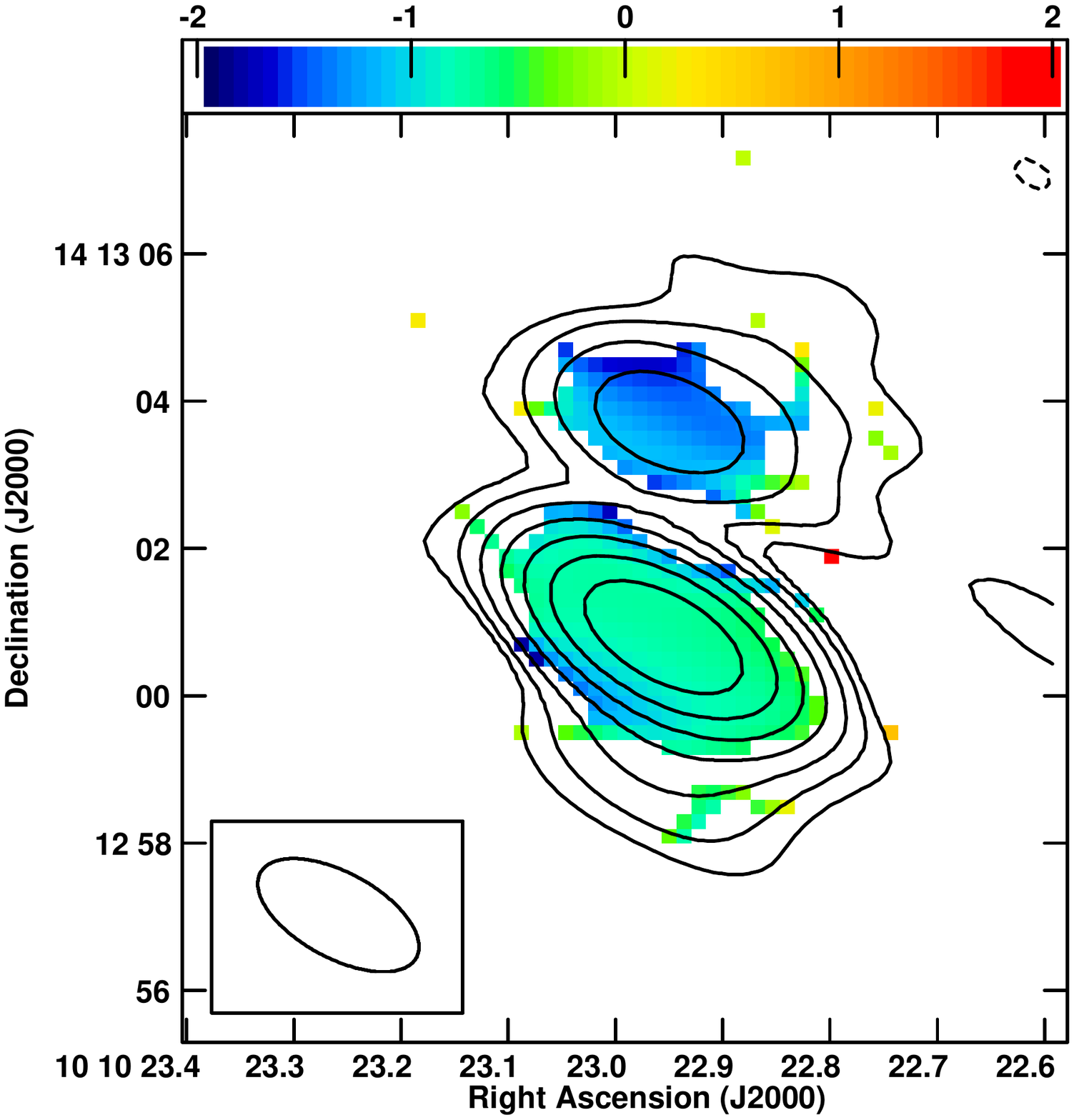}
\includegraphics[width=9.4cm, trim=0 200 0 190]{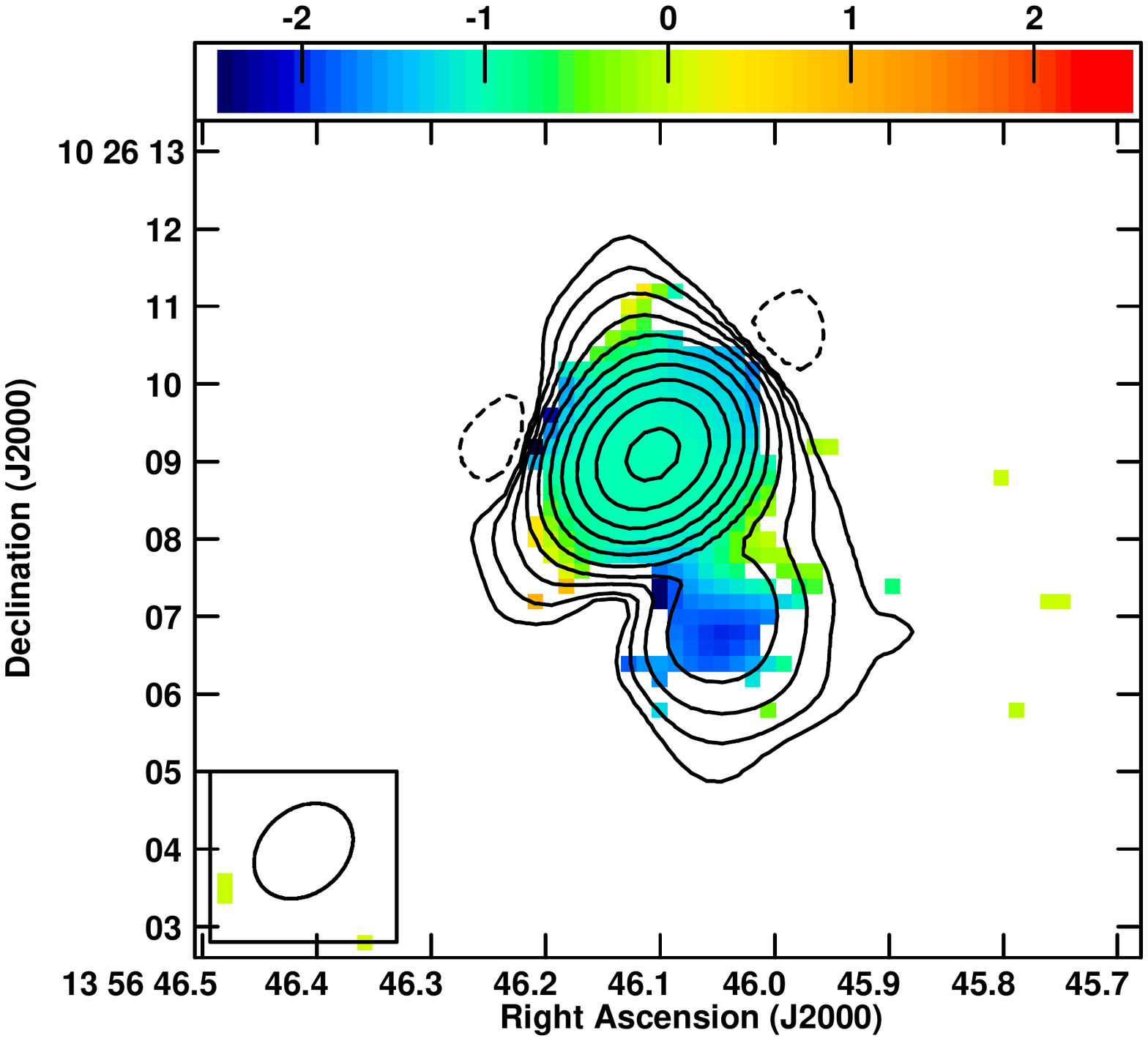}}
\centerline{
\includegraphics[width=10.0cm, trim=0 200 0 200]{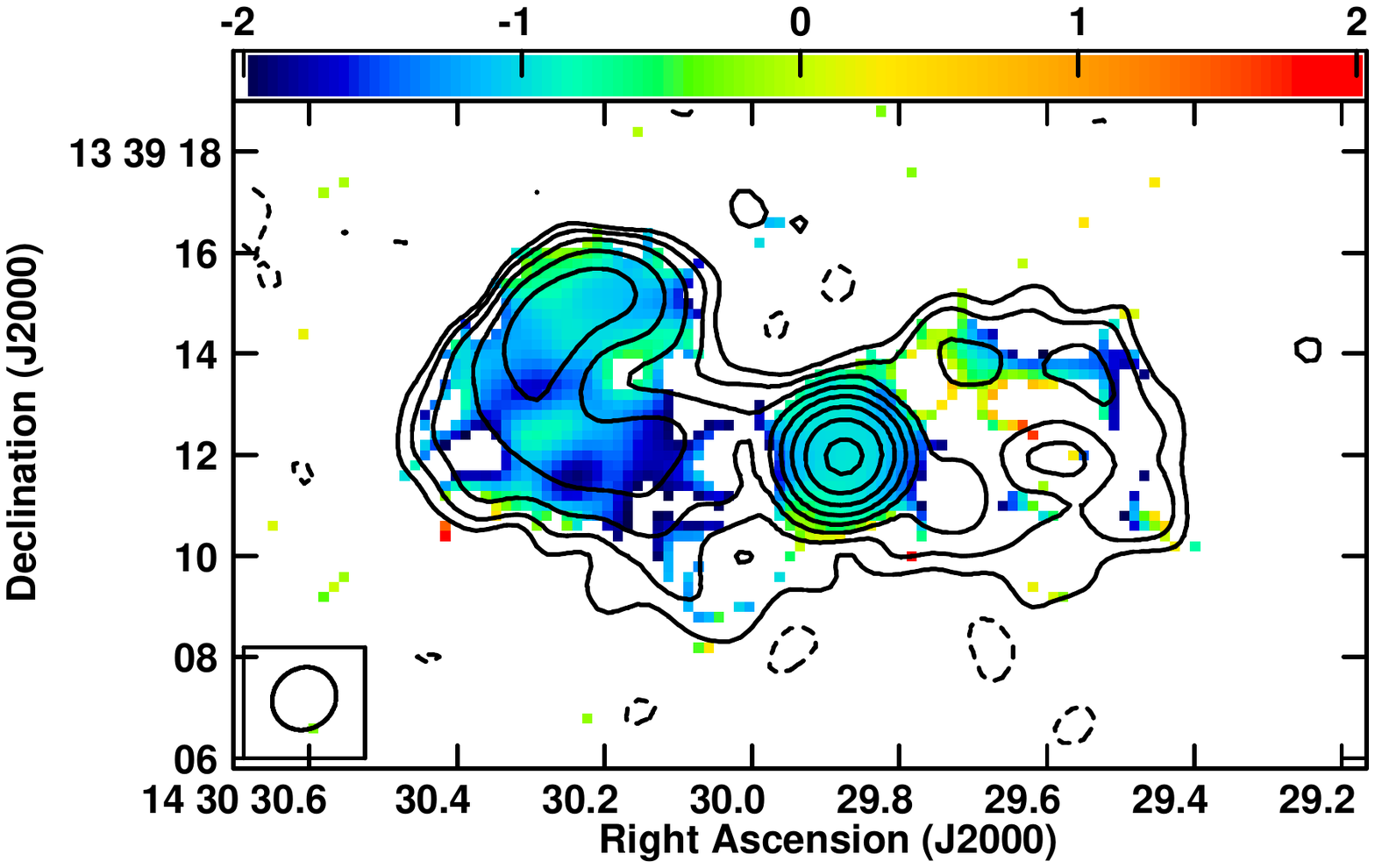}}
\caption{VLA 5 GHz B-array total intensity contours in black with in-band ($\sim$ 4$-$6 GHz) spectral index image in color for (top left) J0945+1737, (top right) J1000+1242, (middle left) J1010+1413, (middle right) J1356+1026 and (bottom) J1430+1339. The color scale ranges from $-$2 to 2 for J0945+1737, J1000+1242, J1010+1413 and J1430+1339, and $-$2.5 to 2.5 for J1356+1026. The peak contour surface brightness is {\it x} mJy beam$^{-1}$ and the levels are {\it y} $\times$ (-1, 1, 2, 4, 8, 16, 32, 64, 128, 256, 512) mJy beam$^{-1}$, where ({\it x} ; {\it y}) is (9.4 ; 0.17), (12 ; 0.0265), (3.8 ; 0.031), (18 ; 0.030) and (3.0 ; 0.019) for J0945+1737, J1000+1242, J1010+1413, J1356+1026 and J1430+1339, respectively.}
\label{figA1}
\end{figure*}

The VLA 5 GHz B-array total intensity contours in black superimposed with in-band ($\sim$4-6 GHz) spectral index in color for individual sources are presented in Figure~\ref{figA1}. The mean spectral index values ($\alpha$) for different radio components estimated from the in-band spectral index images of individual sources are presented in Table~\ref{table3}, Column 4. We also provide spectral index values for different radio components from $\sim$1$\arcsec$ images of \citet{Jarvis19} (their table 4, Column 8) in Table~\ref{table3}, Column 5. They estimated spectral index values by fitting a line through all the detected points at frequencies 1.5, 5.2 and 7.2 GHz. We find that the spectral index values obtained using both methods are broadly consistent with each other, suggesting that the in-band $\alpha$ measurements are a good approximation of the spectral index estimates. 

Overall, we find that $\alpha$ for both radio core and lobes vary from steep ($\alpha<-0.5$) to ultra-steep ($\alpha<-1$). It is interesting to find optically thin cores in these sources. A steep spectrum core either suggests the presence of unresolved jet/lobe emission or optically thin synchrotron emission from AGN/starburst-driven winds \citep[e.g.][]{Hwang18}. Thus, the steep/ultra-steep spectrum cores in our sources except J1356+1026, are consistent with the presence of core-jet structures revealed in their $\sim0.25\arcsec$ images. The $\sim0.25\arcsec$ image of J1356+1026 detects only the radio core (HR:A) with an in-band (4–8 GHz) spectral index of $-0.8$ \citep{Jarvis19}. This suggests that the steep spectrum core in J1356+1026 is either due to the presence of jet/lobe emission on even smaller spatial scales than has been probed by the current radio observations or due to the presence of AGN/starburst-driven winds \citep[e.g.][]{Silpa20}.

Also, the steep/ultra-steep spectrum lobes in our sources are consistent with optically thin synchrotron emission. However, we also note that the ultra-steep spectrum of $\alpha<-1$ is often associated with `relic' emission arising from the past episodes of AGN activity \citep[e.g.][]{Roettiger94, Kharb16}. Moreover, the mis-alignment between the PAs of the jet/lobe structures in the $\sim0.25\arcsec$ and $\sim1\arcsec$ images of our sources, and the differences in their distances to the potential cores (which are taken to be the $\sim0.25\arcsec$ cores), could arise from multiple episodic activities of a precessing jet that changes its direction \citep[e.g.][]{Kharb06, Gallimore06, Orienti16, Jarvis19}.

\bsp	
\label{lastpage}
\end{document}